\documentclass[11pt]{article}
\usepackage[utf8]{inputenc}

\usepackage{graphicx}
\usepackage{amsmath,amssymb}
\allowdisplaybreaks
\usepackage{latexsym}
\usepackage{subfigure}
\usepackage{wrapfig}
\usepackage{xcolor}
\usepackage{float}
\usepackage{hyperref}
\usepackage{comment}
\usepackage{soul}
\usepackage{authblk}
\usepackage{adjustbox}
\usepackage{cases}
\usepackage[titletoc]{appendix}
\usepackage{multirow}
\usepackage{longtable}

\usepackage[left]{lineno}
\usepackage[backend=biber,style=nature]{biblatex}
\date{}
\begin{document}
\title{Lattice Compatibility and Energy Barriers in Intercalation Compounds}
\author[1]{Delin Zhang}
\affil[1]{\small{Aerospace and Mechanical Engineering, University of Southern California, Los Angeles, CA 90089, USA}}
\author[1,2]{Ananya Renuka Balakrishna\thanks{Corresponding author: ananyarb@ucsb.edu}}
\affil[2]{\small{Materials Department, University of California, Santa Barbara, CA 93106, USA}}

\maketitle
\begin{abstract}
\noindent We present a continuum model for symmetry-breaking phase transformations in intercalation compounds, based on Ericksen's multi-well energy formulation. The model predicts the nucleation and growth of crystallographic microstructures in Li$_2$Mn$_2$O$_4$---a representative intercalation compound---with twin boundary orientations and volume fractions that closely match experimental observations. Our chemo-mechanically coupled model not only generates geometrically accurate microstructures through energy minimization, but also reveals a subtle interplay between twinned domains and electro-chemo-mechanical behavior. A key finding is that intercalation compounds satisfying specific compatibility conditions (e.g., $\lambda_2 = 1$ or $|\mathrm{det}\mathbf{U}-1|=0$) show lower elastic energy barriers, require smaller driving forces, and display narrower voltage hysteresis loops. Furthermore, we show that twinned domains act as conduits for fast Li-diffusion. These results establish quantitative design guidelines for intercalation compounds, which focuses on tailoring lattice deformations (rather than suppressing them) and reducing energy barriers to mitigate structural degradation and enhance the electrochemical performance of battery electrodes.
\end{abstract}

\section*{Introduction}
\noindent Intercalation compounds are a class of phase change materials in which guest species, such as ions, atoms, or molecules, are reversibly inserted into host lattices \cite{whittingham1978chemistry}. This reversible insertion makes intercalation compounds ideal for use as electrodes in batteries, supercapacitors, catalysts, and energy harvesters \cite{rajapakse2021intercalation}. Many of these applications require these compounds to be cycled reversibly and repeatedly over a single lifespan. However, inserting a guest species into a host compound commonly induces an abrupt and large deformation of its lattices. This deformation generates significant internal stresses and volume changes in the material, leading to capacity fade, voltage hysteresis, and shortening their lifespans \cite{koerver2017capacity, van2022hysteresis,  santos2023chemistry, lewis2019chemo, renuka2022crystallographic}. To address these issues, current material design strategies focus on minimizing and eliminating changes in lattice geometries during intercalation.

\vspace{2mm}
\noindent Researchers employ advanced techniques to minimize lattice geometry changes, such as site-selective doping of intercalation compounds \cite{zhang2022compositionally, schofield2022doping}, multilayering of heterostructures \cite{xiong2020strain}, pillaring of layered materials \cite{yang2019size}, and designing mechanical constraints (e.g., epitaxial strains, electrode coating, engineering coherent precipitates) \cite{zhang2021film, cho2001zero, wang2022strain}. These zero-strain strategies suppress lattice distortions during intercalation and facilitate the reversible cycling of compounds. For example, the doping efforts by Zhang et al., \cite{zhang2022compositionally} have resulted in a high-entropy Ni-based layered compound that shows nearly zero-volume changes during intercalation, greatly reducing cracking during the charge/discharge cycles. Another intercalation compound Li[Li$_{1/3}$Ti$_{5/3}$]O$_4$, undergoes near zero lattice distortions (i.e., cubic symmetry in both reference ($a_0$ = 8.365 \text{\AA}) and transformed ($a$ = 8.370 \text{\AA}) phases), which in turn contributes to its improved cyclability ($>100$ cycles with 94$\%$ capacity retention) \cite{ohzuku1995zero}.
Other strategies, such as engineering coherent precipitates in NMC cathodes and suppressing structural transformations through a pinning effect, have contributed to improved cyclability \cite{wang2022strain}. Similarly, in our previous work, we showed that epitaxial strains can be designed to suppress phase transformations with large lattice deformations in Li$_x$V$_2$O$_5$ \cite{zhang2021film}. In doing so, we demonstrated reversible structural transformations in Li$_x$V$_2$O$_5$ despite operating in a wide voltage
window \cite{zhang2021film}. These strategies have identified new material compositions and innovative boundary constraints and have initiated a `zero-strain' approach to designing intercalation compounds with improved lifespans \cite{bonnick2018insights, yang2022new, zhao2022zero,zhao2022design}.

\vspace{2mm}
\noindent In contrast to the `zero-strain' design strategy that aims to suppress or eliminate lattice geometry changes, we draw inspiration from ferroelastic materials, in which lattices undergo finite deformations but still generate microstructures with stress-free interfaces and volume-preserving macroscopic changes \cite{bhattacharya2003microstructure, renuka2022crystallographic}. The lattice geometry changes in these ferroelastic materials (e.g., shape-memory alloys) satisfy specific compatibility criteria, deform along defined habit planes, and collectively generate crystallographic microstructures at the continuum scale \cite{ball1987fine}. This relationship between lattice deformations and continuum microstructures is well-established for ferroelastic materials and other functional materials, including ferroelectrics, and ferromagnets \cite{chu1995analysis, renuka2016nanoscale, renuka2022design}. Researchers have extensively used this lattice-continuum link as a design guideline and carefully engineered lattice geometries in phase transformation materials to achieve remarkable reversibility and enhanced material lifespans.
\begin{figure}[t!]
    \centering
    \includegraphics[width=0.8\textwidth]{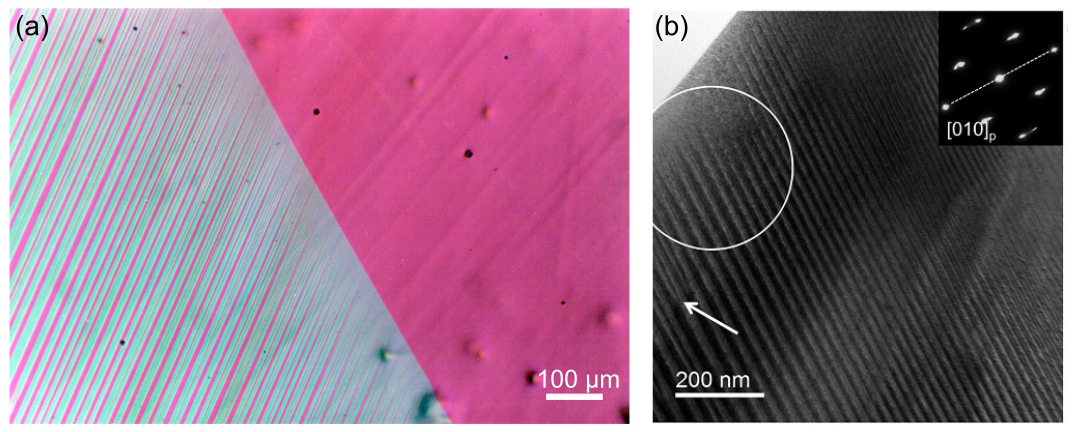}
    \caption{Phase transformation microstructures in shape-memory alloys and intercalation compounds bear similarities in that they generate a mixture of finely twinned domains. (a) An austenite-martensite interface showing the finely twinned martensitic domains in the $\mathrm{Cu}$-14.0\%$\mathrm{Al}$-3.9\%$\mathrm{Ni}$ shape-memory alloy. Image courtesy of Chu and James \cite{chu1995analysis}. (b) A bright field image of a partially transformed Li$_2$Mn$_2$O$_4$ showing the twinned tetragonal variants (Reproduced with permission from American Chemical Society \cite{erichsen2020tracking}).}
    \label{Fig0}
\end{figure}

\vspace{2mm}
\noindent A lattice compatibility criterion known as the middle eigenvalue condition ($\lambda_2 = 1$) has emerged as an important guideline in materials design \cite{chen2013study}. This criterion signifies an `unstretched' edge between the lattices of reference and transformed regions during phase transformation. Materials satisfying this criterion generate two-phase microstructures with phase boundaries that are compatible and stress-free. Theoretically, these interfaces have zero elastic energy and offer minimal energy barriers for phase transformation. Other lattice geometry relationships, such as the cofactor conditions \cite{chen2013study} and self-accommodation (or volume-preserving) conditions \cite{knupfer2013nucleation}, also significantly reduce the elastic energy built in the material during phase transformations. These special lattice deformations have had a significant impact in mitigating structural decay in materials and enhancing reversible phase transformations. For example, shape-memory alloys satisfying the compatibility criteria are shown to have ultra-low fatigue \cite{chluba2015ultralow}, reduced thermal hysteresis \cite{zarnetta2010identification}, and enhanced reversibility \cite{song2013enhanced}. These lattice geometry relations have also informed the design of several functional materials with improved lifespans, such as ferroelectrics, semiconductors, and ferromagnets \cite{wegner2020correlation, liang2020tuning, liu2019giant, balakrishna2022compatible}.

\vspace{2mm}
\noindent Phase transformations in intercalation compounds are analogous to the first-order phase transformations observed in structural and functional materials \cite{renuka2022crystallographic}, see Fig.~\ref{Fig0}. The host lattices in intercalation compounds such as Li$_2$Mn$_2$O$_4$, LiCoO$_2$, and LiFePO$_4$, undergo abrupt and displacive transformations during Li-intercalation \cite{ mizushima1981lixcoo2, padhi1997phospho,luo2020operando,erichsen2020tracking}. The scalar Li-composition field driving phase transformations in intercalation compounds is similar to the thermal field driving phase change in shape-memory alloys. Additionally, intercalation compounds with Jahn-Teller active elements, such as Li$_2$Mn$_2$O$_4$ and NaMnO$_2$, undergo symmetry-breaking lattice deformations characteristic of shape-memory alloys \cite{erichsen2020tracking, luo2020operando, abakumov2014multiple}. The symmetry-breaking transformations in intercalation compounds generate finely twinned microstructures that are quantitatively consistent with the crystallographic theory of martensites (see Fig.~\ref{Fig0} and Refs.~\cite{zhang2023designing, zhang2024coupling, chu1995analysis}). While there are a few exceptions between intercalation compounds and ferroelastics, such as the anisotropy of Li-diffusion in select compounds \cite{tang2011anisotropic} and cation migration or dissolution in disordered cathodes \cite{clement2020cation, choa1999structural}, we will initially focus on intercalation compounds in which the host lattices undergo exclusively displacive structural transformations.

\vspace{2mm}
\noindent The crystallographic microstructures not only minimize the elastic energy during phase change but also are reported to alter the ion-transport properties in intercalation compounds. For example, a first-principles study in LiCoO$_2$ demonstrates faster diffusion of lithium along twin boundaries when compared to Li-diffusion across twin interfaces \cite{moriwake2013first}. Similar coherent interfaces with low interfacial energy have been reported in other intercalation compounds such as NaMnO$_2$  \cite{abakumov2014multiple}. These twinned microstructures were imaged in large concentrations and were found to be thermodynamically stable at equilibrium voltages \cite{abakumov2014multiple}. These findings reveal the common occurrence and stability of twin boundaries in intercalation compounds and highlight the potential for crystallographically designing twin interfaces to serve as conduits for fast charge/discharge conditions. 

\vspace{2mm}
\noindent More recently, we developed algorithms to screen intercalation compounds ($n > 5,000$ pairs from open source material databases) that undergo displacive phase transformations and identified candidate materials that approximately satisfy the lattice compatibility criteria for stress-free interfaces and volume-preserving microstructures \cite{zhang2023designing}. Our results showed that intercalation compounds with spinel and NASICON structures typically undergo symmetry-breaking lattice deformations and are suitable to generate energy-minimizing crystallographic microstructures \cite{zhang2023designing, erichsen2020tracking, shin2004factors}. 
Other symmetry-breaking transformations are reported in Prussian blue analogues \cite{wang2020reversible,zhang2022lithiated} during
alkali-ion (e.g., Na$^+$ and Li$^{+}$) insertion/extraction.
To predict these crystallographic microstructures, we develop a multi-variant continuum model using an energy landscape that is informed by lattice symmetry with Li$_2$Mn$_2$O$_4$ as a representative example \cite{zhang2024coupling}. A key feature of our model is that we not only distinguish between the phases in terms of Li-compositions (e.g., Li-rich or Li-poor), but also differentiate between the various tetragonal lattice variants formed during the symmetry-breaking transformation. By minimizing the total energy across a multi-well energy landscape, we demonstrate the nucleation and growth of crystallographic microstructures in an electrochemical environment. 

\vspace{2mm}
\noindent  In this work, we propose a material design approach that emphasizes \textit{compatibility}, i.e., the fitting together of phases during transformation. Using our multi-variant continuum model, we predict microstructures in Li$_2$Mn$_2$O$_4$ that are geometrically accurate, consistent with experimental observations, and arise naturally from elastic energy minimization. We demonstrate that allowing certain lattice deformations in intercalation compounds---provided they satisfy specific compatibility criteria---can minimize internal stresses and net volume changes. In this approach, lattice deformations are carefully engineered to mitigate interfacial stresses at phase boundaries and to accommodate volume changes that accompany phase change in intercalation compounds. In doing so, we show how precise lattice geometries can substantially lower the elastic energy barriers that emerge during phase transformation, thereby reducing voltage hysteresis and enhancing material reversibility. These energy barriers and energy-minimizing microstructures, characteristic of first-order phase transformations, are difficult to estimate using first-principles calculations, but are effectively captured by our continuum approach. Furthermore, we find that crystallographic microstructures facilitate faster Li-diffusion along twin boundaries and can be intentionally designed in intercalation compounds to enable rapid charge/discharge cycles. Our theoretical results reveal lattice geometry relationships---beyond the conventional `zero-strain' condition---that provide new guidelines for designing intercalation compounds with improved lifespans.

\section*{Results}
\subsection*{Continuum Theory of Crystalline Solids}
The Cauchy-Born rule states that the overall deformation gradient of the macroscopic body maps onto the deformation gradient of individual lattices at the atomic scale. This lattice-continuum link plays an important role in predicting microstructures that emerge during phase change \cite{ericksen2008cauchy} and in the crystallographic designing of materials with enhanced lifespans. The deformation gradient can be decomposed into a unique rotation and a stretch tensor $\mathbf{F = QU}$. The stretch tensor $\mathbf{U}$ is a positive-definite symmetric matrix and describes the displacive distortions of the lattices between the reference and transformed phases. We construct this stretch tensor using lattice geometries of the crystalline solid, and it is an important input to predict the geometric features of microstructures, such as the orientation of twin interfaces and phase boundaries.
\begin{figure}[ht!]
    \centering
    \includegraphics[width=0.8\textwidth]{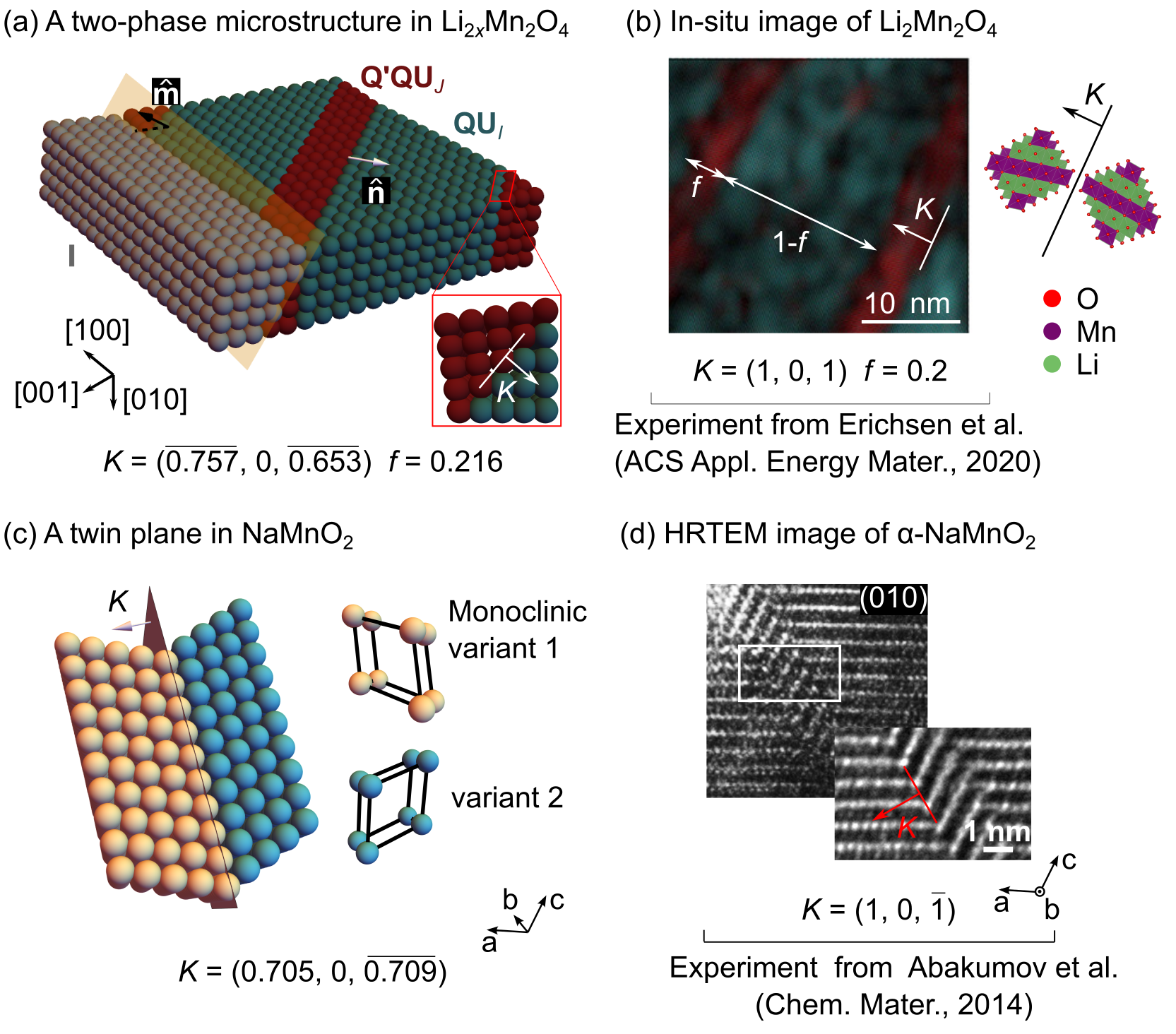}
    \caption{A comparison between theoretical and experimental microstructures in representative intercalation compounds. Using the lattice geometries of Li$_2$Mn$_2$O$_4$ and NaMnO$_2$ as inputs, we analytically derive the twin solutions as shown in Tables~S2 and S3. We use these solutions to geometrically construct the microstructures shown in (a) and (c), respectively. (a) The twinned microstructure in spinel Li$_{2x}$Mn$_2$O$_4$, with vectors $\mathbf{\hat{m}}$ and $\mathbf{\hat{n}}$ representing the orientations of the phase boundary and the twin plane, respectively. Our calculations predict a volume fraction of the twinned mixture to be $f = 0.216$, in agreement with experimental measurements \cite{erichsen2020tracking}. (b) Bragg-filtered HRTEM image of the lamellar microstructures from Erichsen et al. \cite{erichsen2020tracking}, showing a twin plane of (1,~0,~1) orientation and a volume fraction of $f = 0.2$. (c) Geometric construction of a twin interface between two monoclinic variants of NaMnO$_2$, with an orientation given by $K = (0.705,~0,~\overline{0.709})$. (d) Our geometric construction matches previously reported HRTEM imaging of a NaMnO$_2$ sample by Abakumov et al. \cite{abakumov2014multiple}, which shows the $(1,~0,~\bar{1})$ direction of the twins in the monoclinic phase. (Reprinted with permission from American Chemical Society \cite{abakumov2014multiple}).}
    \label{Fig14}
\end{figure}

\vspace{2mm}
\noindent Intercalation compounds that undergo a symmetry-breaking transformation generate continuum microstructures with fine plate-like crystallographic features, see Fig.~\ref{Fig14}(b). We show that the twinned microstructures in intercalation compounds form as a consequence of elastic energy minimization and satisfy the kinematic compatibility conditions applicable to all ferroelastic materials \cite{ball1987fine}:
\begin{gather}
\mathbf{Q} \mathbf{U}_J - \mathbf{U}_I = \mathbf{a} \otimes \hat{\mathbf{n}}, \label{eq:twin equation} \\
\mathbf{Q}' \left( f \mathbf{Q} \mathbf{U}_J + (1 - f) \mathbf{U}_I \right) = \mathbf{I} + \mathbf{b} \otimes \hat{\mathbf{m}}. \label{eq:AM interface}
\end{gather}

\noindent Eqs.~(\ref{eq:twin equation}) and (\ref{eq:AM interface}) describe the geometric features of the crystallographic microstructure formed by a mixture of lattice variants (with stretch tensors, $\mathbf{U}_I, \mathbf{U}_J$). The solutions to these equations, determine the orientation of twin planes (vectors $\mathbf{a}, \hat{\mathbf{n}}$), phase boundaries (vectors $\mathbf{b}, \hat{\mathbf{m}}$), and the volume fraction $f$ of twinned domains. 

\vspace{2mm}
\noindent Fig.~\ref{Fig14}(a-d) shows the theoretical and experimental comparison of the twinned domains in two representative intercalation compounds, namely Li$_2$Mn$_2$O$_4$ and NaMnO$_2$. These compounds undergo a symmetry-breaking transformation on intercalation, i.e., cubic-to-tetragonal in Li$_2$Mn$_2$O$_4$ and orthorhombic-to-monoclinic in NaMnO$_2$, which generates multiple lattice variants. Using the stretch tensor for Li$_2$Mn$_2$O$_4$ (see Fig.~\ref{Fig14}(a-b)), we calculate a volume fraction of $f = 0.216$ for the twinned domains and analytically derive the normal vector of the twin plane to be $K=\frac{\mathbf{U}_I^{-1}\hat{\mathbf{n}}}{|\mathbf{U}_I^{-1}\hat{\mathbf{n}}|}=(\overline{0.757},\ 0,\ \overline{0.653})$. These theoretical results align favorably with the experimental Bragg-filtered HRTEM measurements of the twin plane orientation (1,~0,~1) and the volume fraction $f=0.2$ in Ref.~\cite{erichsen2020tracking}. Similarly, using the stretch tensor for NaMnO$_2$ (see Fig.~\ref{Fig14}(c-d)), we geometrically construct a twin interface between two monoclinic variants of NaMnO$_2$. We calculate a twin plane orientation of $K = (0.705,~0,~ \overline{0.709})$ that matches the previously reported $(1,~0,~\bar{1})$ direction shown in the HRTEM image of NaMnO$_2$ by Abakumov et al. \cite{abakumov2014multiple}. Overall, we find that the geometric features imaged in intercalation compounds (e.g., twin plane, volume fraction) compare well with experimental measurements and support the use of the Cauchy-Born rule and the compatibility conditions to predict microstructures in intercalation compounds.
\begin{figure}[ht!]
    \centering
    \includegraphics[width=\textwidth]{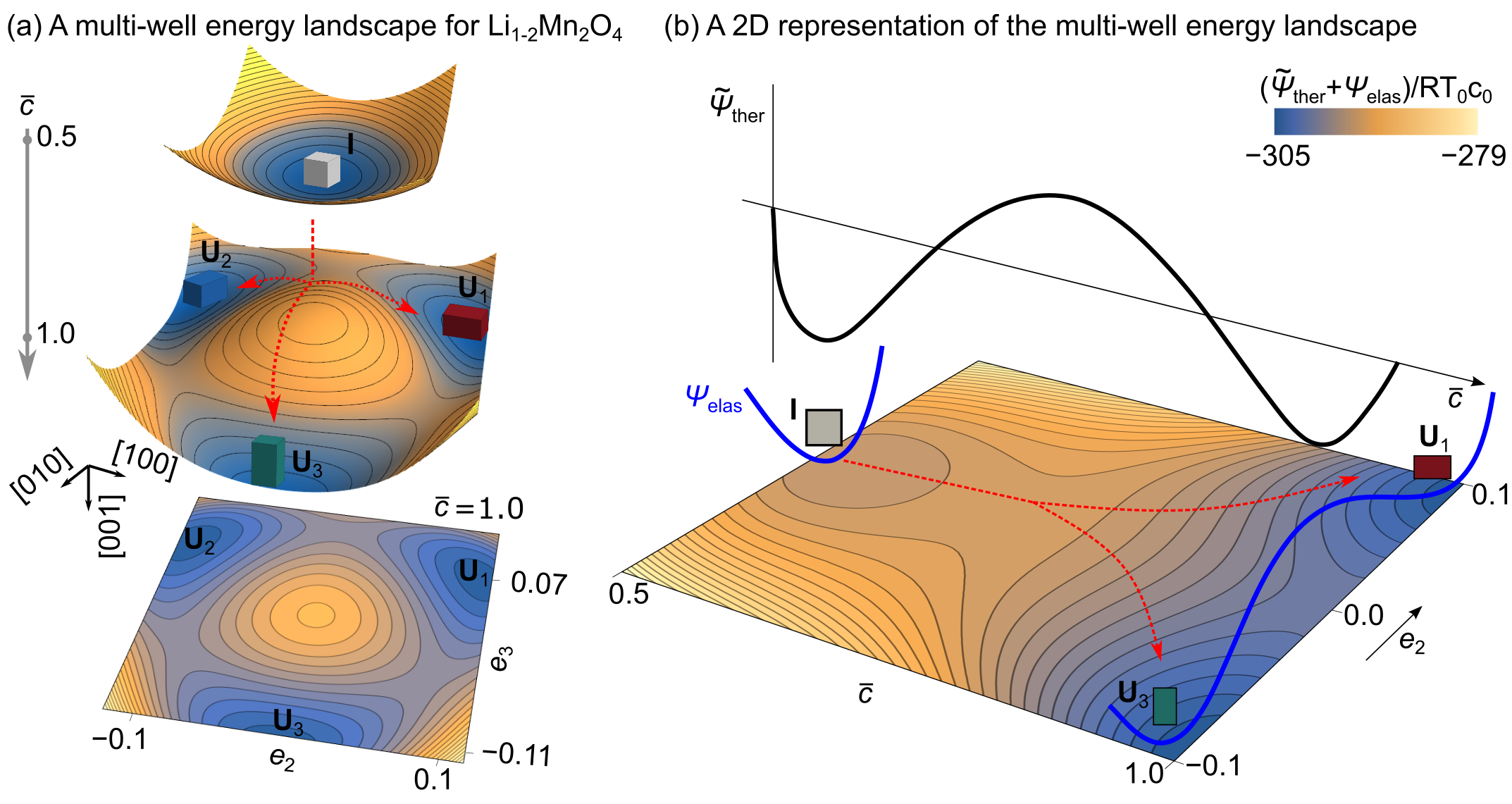}
    \caption{The multi-well energy landscape describing the symmetry-breaking phase transformation in Li$_{2x}$Mn$_2$O$_4$ (0.5 $\leq x \leq$ 1). (a) Contour plots of the free energy as a function of the order parameters: Li-composition $\bar{c}$ and strain variants $(e\mathrm{_2},~e_{\mathrm{3}})$. The single well at $\bar{c} = 0.5$ corresponds to the high-symmetry cubic phase (LiMn$_2$O$_4$), and the three equivalent energy wells at $\bar{c} = 1.0$ correspond to the three tetragonal variants of Li$_2$Mn$_2$O$_4$. (b) In 2D, we represent the cubic-to-tetragonal transformation with a square unit cell for the reference phase at $(\bar c,~e_2) = (0.5,~0)$ and rectangular unit cells for the transformed phase at $(\bar c,~e_2) = (1.0,~\pm 0.1)$. The double-well potential describes the thermodynamic energy $\tilde\psi_{\mathrm{ther}}$, with minima at $\bar{c}=0.5$ and $\bar{c}=1.0$, respectively.}
    \label{Fig4}
\end{figure}

\subsection*{Geometric Microstructures in Li$_2$Mn$_2$O$_4$}
\label{sec:Geometric Microstructures in LMO}
\noindent We construct a multi-well free-energy landscape for Li$_{2x}$Mn$_2$O$_4$ as a function of Li-composition $c$ and host-lattice deformation gradient $\mathbf{F=QU}$. Inserting Li into a cubic LiMn$_2$O$_4$ induces a Jahn-Teller lattice distortion, which generates three tetragonal variants of Li$_2$Mn$_2$O$_4$. We define two order parameters: the normalized Li-composition $\bar{c}$ and a strain measurement vector $\mathbf{e} = \{e_1, \ e_2,\ e_3,\ \dots, \ e_6\}^\top$. The Li-composition distinguishes between the LiMn$_2$O$_4$ and Li$_2$Mn$_2$O$_4$ phases, while strain components (e.g., $e_2$, $e_3$) distinguish between Li$_2$Mn$_2$O$_4$ lattice variants, see Fig.~\ref{Fig4}(a). The explicit form of the free energy function and the governing equations are described in the Methods section.

\begin{figure}[ht!]
    \centering
    \includegraphics[width=0.6\textwidth]{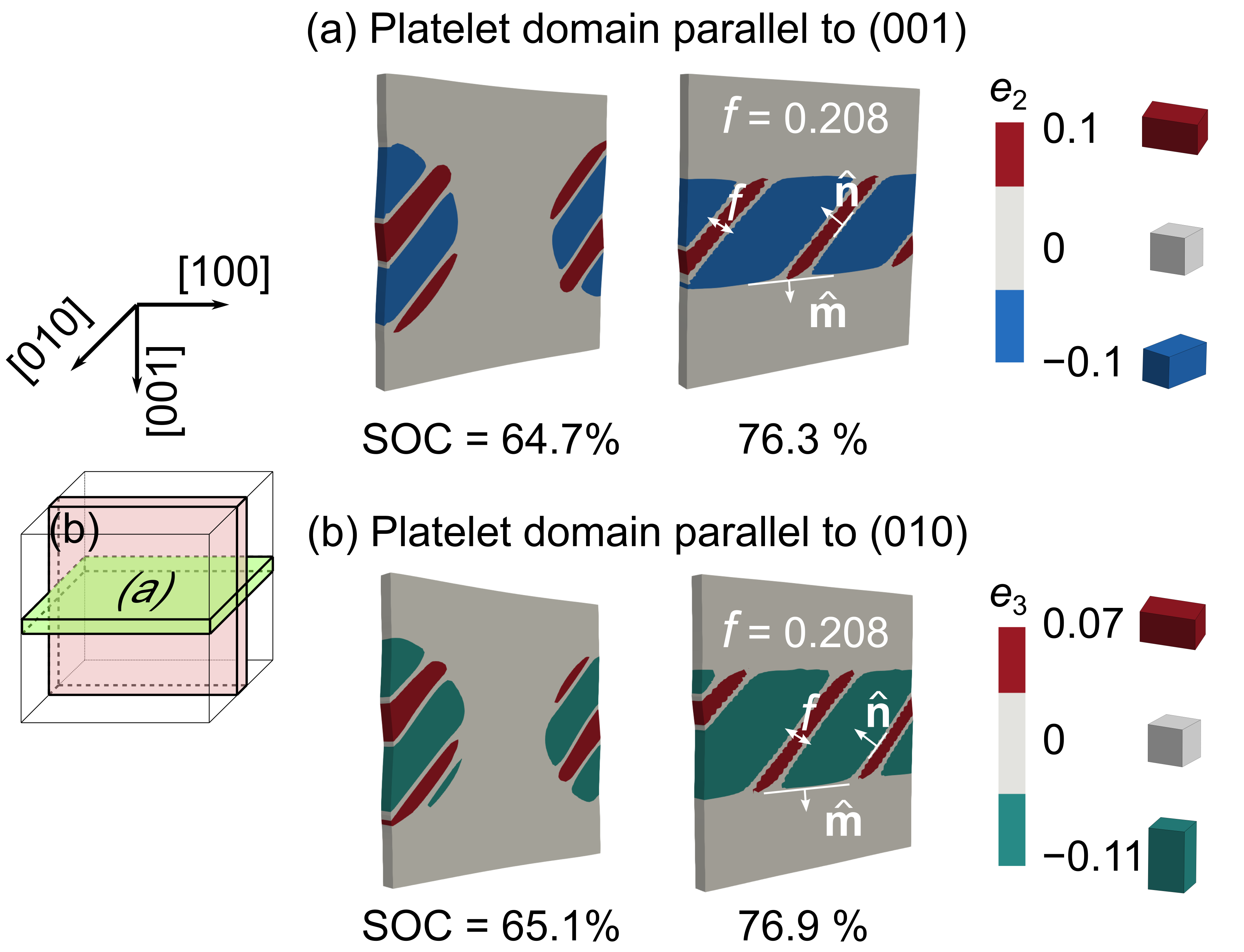}
    \caption{The nucleation and growth of twinned microstructures in Li$_{2x}$Mn$_2$O$_4$ on platelet-shaped 3D computational domains. In both (a) and (b), our multi-variant continuum model predicts crystallographic microstructures with a volume fraction $f$ = 0.208 for the twinned domains. These results are consistent with the bright-field imaging of Li$_{2x}$Mn$_2$O$_4$, which shows $f$ = 0.2 \cite{erichsen2020tracking}, and with our analytical calculations in Table~S2.}
    \label{Fig13}
\end{figure}

\vspace{2mm}
\noindent Fig.~\ref{Fig13} shows the nucleation and growth of crystallographic microstructures in Li$_2$Mn$_2$O$_4$ during a discharge half cycle at 0.5C-rate in 3D. Lithiating the reference phase LiMn$_2$O$_4$ induces an abrupt cubic-to-tetragonal transformation of the host lattices. The order parameters $e_2$ and $e_3$ collectively describe the lattice deformation and the three tetragonal variants $\mathbf{U}_1, \mathbf{U}_2, \mathbf{U}_3$ are positioned equidistant from the reference energy well $\mathbf{I}$, see Fig.~\ref{Fig4}(a). The lattice misfit between the cubic phase and any single variant of the tetragonal phase is significant $\sim 13\%$ and contributes to the elastic energy at the phase boundary. To minimize this energy, the tetragonal variants form a fine mixture (e.g., $\mathbf{U}_1$ and $\mathbf{U}_2$ in Fig.~\ref{Fig13}(a)) with an average deformation that is compatible with the cubic phase. These finely twinned microstructures (also referred to as the classic austenite/martensite microstructure) minimize coherency stresses at the phase boundary, and are formed as a consequence of elastic energy minimization.

\begin{figure}[t!]
    \centering
    \includegraphics[width=\textwidth]{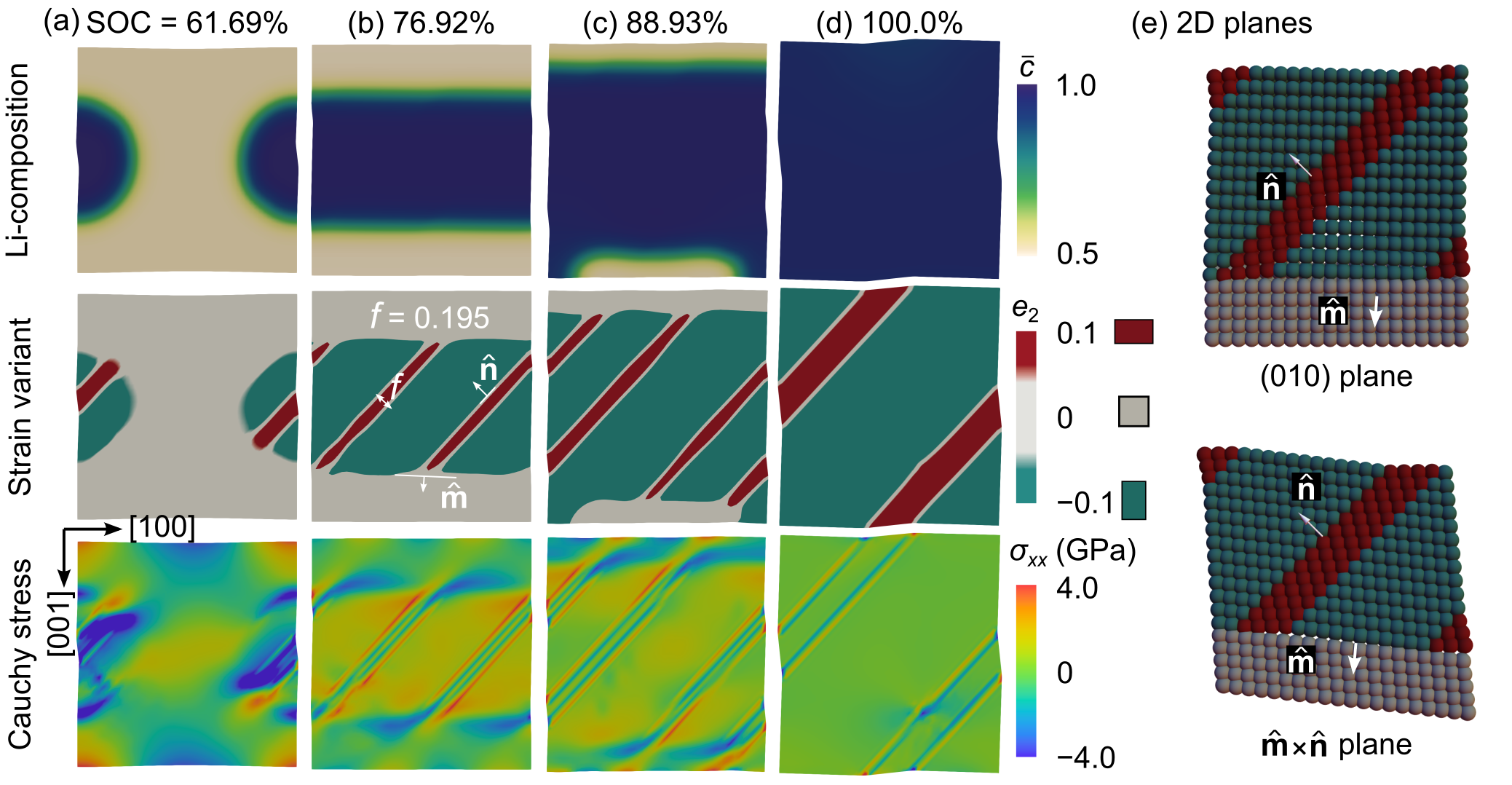}
    \caption{The nucleation and growth of twinned microstructures in Li$_{2x}$Mn$_2$O$_4$ during a discharge half cycle at 0.5C-rate. (a-d) The images on the top and middle rows, respectively, illustrate the evolution of Li-composition $\bar c$, and the strain variant $e_2$ as a function of the SOC. The geometric features of the twinned microstructures, such as the twin plane orientation $\hat{\mathbf{n}}$, volume fraction $f$, and the approximate orientation of the phase boundary $\hat{\mathbf{m}}$, are highlighted. The images on the bottom row show the Cauchy stresses $\sigma_{xx}$ with compressive stresses of up to $-4$ GPa across the phase boundary. (e) We project microstructures onto the (010) and $\hat{\mathbf{m}} \times \hat{\mathbf{n}}$ planes, respectively. The twin plane orientations and the volume fraction are geometrically accurate on the $\hat{\mathbf{m}} \times \hat{\mathbf{n}}$ plane; however, have minor errors ($\pm5\%$) on the (0,~1,~0) plane because of negligible out-of-plane distortions in Li$_2$Mn$_2$O$_4$.}
    \label{Fig5}
\end{figure}

\vspace{2mm}
\noindent Fig.~\ref{Fig13}(a-b) shows two examples of crystallographic microstructures that evolve in platelet-shaped 3D computational domains. These calculations predict the volume fraction of the twinned mixture as $f=0.208$ and a twin plane orientation of $45^\circ$. These results compare favorably with the experiments by Erichsen et al. \cite{erichsen2020tracking}, in which bright field imaging of Li$_2$Mn$_2$O$_4$ microstructure shows a volume fraction of $f=0.2$ and twin orientation of $45^\circ$. These crystallographic microstructures are necessarily three-dimensional; however, their geometric features can be accurately captured on specific 2D planes (e.g., $\hat{\mathbf{m}} \times \hat{\mathbf{n}}$ in Fig.~\ref{Fig5}(e)). This dimensionality reduction would benefit numerical calculations, which are expensive in 3D because of the higher-order PDEs involved and the challenges associated with stabilizing finite element calculations (see the Methods section). For Li$_2$Mn$_2$O$_4$, the out-of-plane strain component is small ($\frac{a-a_0}{a_0}=-0.03$) when compared to its in-plane strain component ($\frac{c-c_0}{c_0}=0.13$). This small out-of-plane distortion explains the $\sim 2\%$ error in twin plane orientation ($\Delta\theta \approx 0.89^\circ$) and volume fractions ($\Delta f \approx 0.01$) reported in experiments (i.e., Erichsen et al. \cite{erichsen2020tracking} imaged Li$_2$Mn$_2$O$_4$ microstructures on the (010) plane). With these differences in mind, we proceed to characterize and investigate these on a representative 2D plane.

\vspace{2mm}
\noindent Fig.~\ref{Fig5}(a-d) shows the nucleation and growth of the 2D Li$_2$Mn$_2$O$_4$ microstructures with increasing state-of-charge (SOC = $\int_\Omega\bar{c}dV/V$). The geometric features of the twinned domains satisfy kinematic compatibility criteria, and the volume fraction of twins, twin plane orientation, and phase boundaries emerge phenomenologically (without any a priori assumptions) in our computations. In Fig.~\ref{Fig5}(b), the volume fraction of the martensite twins in the Li$_2$Mn$_2$O$_4$ phase is $f \approx 0.2$, which closely matches our analytical solution of $f=0.216$ for Li$_2$Mn$_2$O$_4$ in Eq.~(\ref{eq:AM interface}) and is within $\sim 4\%$ error of our 3D calculations. The twin interfaces are oriented at about 45$^\circ$, which is consistent with the solution of the twinning equation with $\hat{\mathbf{n}} = (\overline{0.707},\ 0,\ \overline{0.707})$. The volume fraction and orientation of the twinned domains are a consequence of the averaging of tetragonal variants (to maintain compatibility with the reference phase) and emerge in the early stages of phase transformation. The microstructural evolution in Fig.~\ref{Fig5} was computed at 0.5C-rate, which is sufficiently slow for instantaneous relaxation of the elastic energy. This explains why the volume fraction and twin boundary orientation correspond to the energy-minimizing solutions in the initial stage of the discharge cycle.

\vspace{2mm}
\noindent In Fig.~\ref{Fig5}(b), the phase boundary separating the uniform LiMn$_2$O$_4$ and the finely twinned Li$_2$Mn$_2$O$_4$ is planar with orientation (0, 0, 1). This orientation of the phase boundary closely matches the analytical solution of the austenite/martensite interface in Fig.~\ref{Fig14}(a). With continued lithiation, the phase boundary propagates through the computational domain. This phase boundary motion is driven not only by elastic energy minimization, but also by the anisotropic Li-diffusion (see next subsection) and surface boundary conditions. Consequently, the orientations of the phase boundary in Fig.~\ref{Fig5}(c) form in a collective response to energy minimization and the applied flux boundary condition.

\vspace{2mm}
\noindent Fig.~\ref{Fig5}(bottom row) shows the Cauchy stress distribution during phase change. Although twin boundaries are exactly compatible as predicted by the sharp interface theory \cite{ball1987fine}, our simulation shows finite stresses at these interfaces. In our multi-variant continuum model, based on the diffuse-interface theory, we penalize changes in the deformation gradient (i.e., $\nabla\mathbf{e}$). This penalty manifests as higher-order stresses at the twin interfaces. Additionally, we note that the interfacial stress at the phase boundary, although minimized because of fine twinning, is not zero. These stresses $\sigma_{xx} \approx 4 \ \mathrm{GPa}$ accumulate with repeated cycling and nucleate microcracks in the material \cite{luo2020operando}. The cubic-to-tetragonal transformation in Li$_2$Mn$_2$O$_4$ is not a volume-preserving deformation (i.e., $|\mathrm{det}\mathbf{U}-1| = 6.4\%$), which in turn builds interfacial stresses between neighboring electrode particles or between an electrode and solid-electrolyte interface. These chemo-mechanical challenges accelerate the decay of intercalation compounds.

\subsubsection*{Anisotropic Diffusion along Twin Interfaces}
\label{sec:Anisotropic Diffusion along Twin Interfaces}
\noindent The finite stresses along twin interfaces and phase boundaries act as additional driving forces for Li-diffusion. These driving forces are non-homogeneous on the computational domain and introduce anisotropy in the evolution of Li-composition in electrode particles. For example, in Eq.~(\ref{eq:free energy}) of the Methods section, the moduli $\mathrm{C} =(c_{11}-c_{12})/2$, $\mathrm{K} =(c_{11}+c_{12})/2$, and $\mathrm{G} =c_{44}$ are linear combinations of the elastic stiffness components $c_{11}$, $c_{12}$, and $c_{44}$ of LiMn$_2$O$_4$. The term related to deviatoric modulus $\mathrm{C}(\frac{\bar{c}-0.75}{0.5-0.75}e_2^2)$ and the term related to bulk modulus $\mathrm{K}(e_1 - \frac{\bar{c}-0.5}{1.0-0.5}\Delta \mathrm{V})^2$ are functions of the chemical composition and affect the stress chemical potential as follows:
\begin{align}
\mu_{\mathrm{elas}}= \frac{1}{c_0}\frac{\partial \psi_{\mathrm{elas}}}{\partial \bar{c}}=-\frac{4}{c_0}\mathrm{C} e^2_2-\frac{4\Delta \mathrm{V}}{c_0}\mathrm{K}(e_1 - \frac{\bar{c}-0.5}{1.0-0.5}\Delta \mathrm{V}),
\end{align}

\noindent The driving force for diffusion (arising from the elastic energy component) is given by:\footnote{The total chemical potential---arising from elastic, thermodynamic, and gradient energy terms---is given by Eq.~(\ref{eq:total chemical potential}).}
\begin{align}
\nabla \mu_{\mathrm{elas}}= -\frac{8}{c_0}\mathrm{C} e_2 \nabla e_{2}-\frac{4\Delta \mathrm{V}}{c_0}\mathrm{K}\nabla e_1+\frac{8(\Delta \mathrm{V})^2}{c_0}\mathrm{K}\nabla \bar{c}.
\label{eq:elastic driving force}
\end{align}

\begin{figure}[t!]
    \centering
    \includegraphics[width=0.9\textwidth]{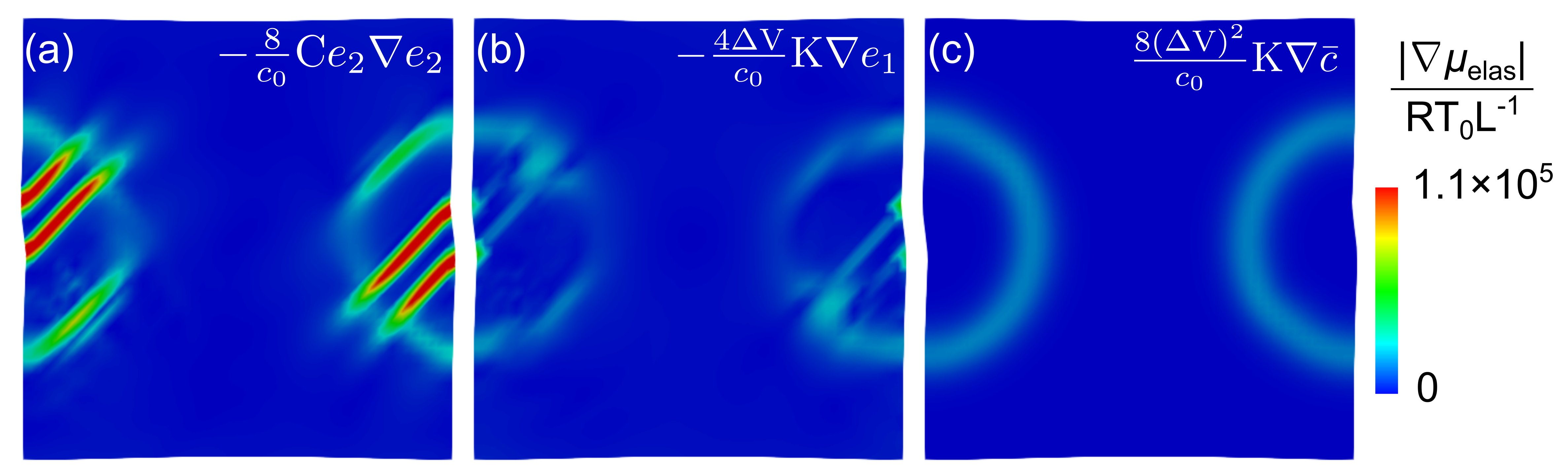}
    \caption{Anisotropic driving forces for Li-diffusion arising from the twinned domains in Li$_2$Mn$_2$O$_4$. (a) The gradient term with $\nabla e_2$ in Eq.~(\ref{eq:elastic driving force}) facilitates faster Li-diffusion along twin boundaries. (b) The effect of $\nabla e_1$ on Li-diffusion is negligible when the dilatational changes to the lattice geometry are small. (c) The gradient of composition $\nabla \bar{c}$ drives Li-diffusion across the phase boundary. Overall, the driving force arising from $-\frac{8}{c_0}\mathrm{C} e_2 \nabla e_{2}$ term is dominant and is primarily present along the twin boundaries; This driving force is absent across the twins.}
    \label{Fig6}
\end{figure}

\noindent Fig.~\ref{Fig6} shows the driving force contributions from individual terms in Eq.~(\ref{eq:elastic driving force}). We note that these driving forces are primarily active along the twin boundaries (arising from $\nabla e_2$ term) but not across the twins. Likewise, the contributions in Fig.~\ref{Fig6}(b-c) are finite along, but not across, the phase boundary (arising from $\nabla \bar{c}$ and $\nabla e_1$ terms). These stress-induced contributions to the chemical potential are in addition to the thermodynamic and gradient energy contributions described in Eq.~(\ref{eq:free energy}).

\vspace{2mm}
\noindent The driving forces in Eq.~(\ref{eq:elastic driving force}) arise from the crystallographic features in Li$_2$Mn$_2$O$_4$ and enhance Li-diffusion along twin and phase boundaries during microstructural evolution. In the absence of experimental measurements of kinetic parameters along twin interfaces or phase boundaries, we use isotropic coefficients for Li-mobility. However, the gradients in the order parameters generate anisotropy in the chemical potential, which facilitates relatively faster Li-diffusion along the twin interfaces when compared to Li-diffusion in the bulk. These results align with evidence that twin boundaries are shown to act as conduits for rapid Li-transport \cite{moriwake2013first}. Overall, Fig.~\ref{Fig6} highlights another potential of engineering crystallographic microstructures in intercalation compounds to facilitate the rapid charging/discharging of electrodes during phase transformation. 

\vspace{2mm}
\noindent Our multi-variant continuum model effectively predicts the crystallographic microstructures that nucleate and grow in Li$_2$Mn$_2$O$_4$. Using lattice geometries of this intercalation compound as the primary input, our continuum model predicts geometrically accurate microstructures that form during electrochemical cycling. These results are consistent with the existing experimental literature \cite{erichsen2020tracking}, and build on these reports by providing quantitative insights into the stress and its contributions in accelerating phase transformation kinetics. The predictive capabilities of the model offer fundamental insights into the microstructural effects on material properties, and open up new possibilities to crystallographically design intercalation compounds. 

\subsection*{Crystallographic Designing}
\label{sec:Crystallographic Designing}
\noindent We use our multi-variant continuum model as a crystallographic design tool to engineer special microstructures with stress-free interfaces $\lambda_2 = 1$ and volume-preserving deformations $|\mathrm{det}\mathbf{U}-1|=0$ during phase change. These microstructures would help address common structural failure issues such as cracking and delamination, which limit the lifetime of solid-state batteries. It is important to note that both these microstructures are more general than the zero-strain condition, in that they allow individual lattices to distort and/or undergo symmetry change during phase transformation, but satisfy specific compatibility criteria. %by doing so, we demonstrate a proof-of-concept for designing materials with enhanced lifespans beyond the zero-strain criterion. 

\begin{figure}[t]
    \centering
    \includegraphics[width=\textwidth]{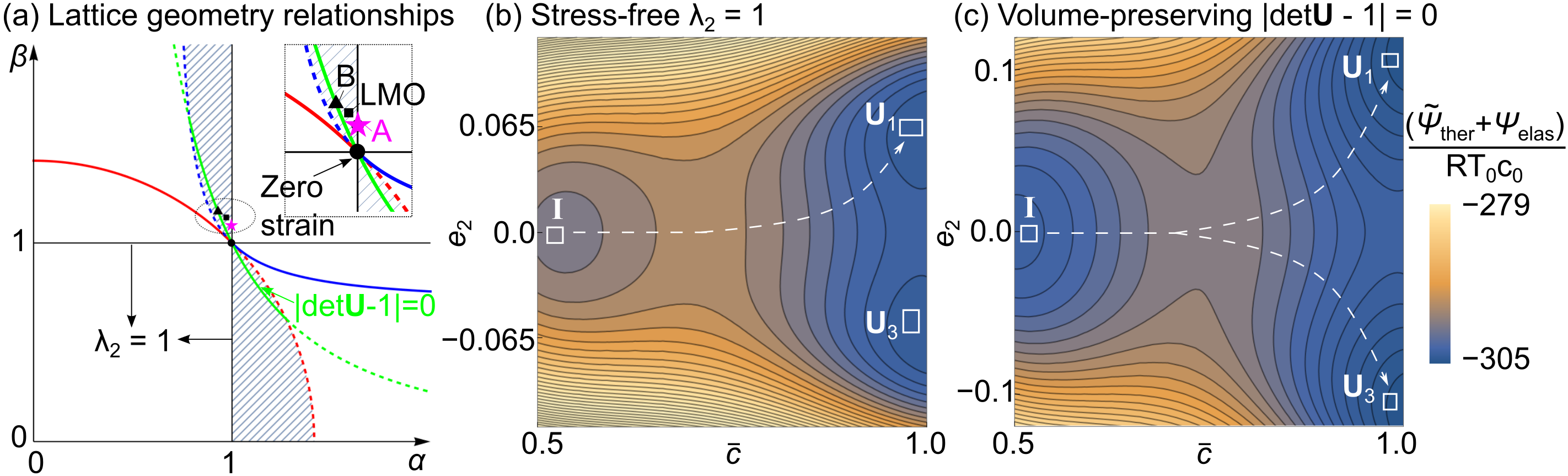}
    \caption{(a) Lattice geometry relationships to form crystallographic microstructures in materials undergoing a cubic-to-tetragonal ($\mathbf{I}\to\mathbf{U}$) symmetry-breaking transformation. The stretch tensor $\mathbf{U}$ is a diagonal matrix with two independent parameters $\alpha = a/a_0$ and $\beta = c/a_0$. The hatched regions show lattice geometries necessary to form the characteristic austenite/martensite interface; the solid-green line inside the hatched region (see inset) corresponds to the volume-preserving ($|\mathrm{det}\mathbf{U}-1|=0$) deformation. The vertical black lines correspond to the $\lambda_2=1$ criterion necessary for stress-free interfaces. The stretch tensor of Li$_2$Mn$_2$O$_4$ is highlighted by `LMO' and we identify two representative lattice geometry relations `A' and `B' that, respectively, satisfy the $\lambda_2 =1$ and $|\mathrm{det}\mathbf{U}-1|=0$ criteria. The `zero-strain' criterion is satisfied only when $\alpha = \beta = 1$. (b) Using the representative lattice geometries at `A', we derive a multi-well energy landscape as a function of Li-composition $\bar{c}$ and strain order parameter $e_2$. By minimizing the total free energy across this landscape, stress-free interfaces emerge as the energy minimizer during phase change. (c) Similarly, using the lattice geometry relation at `B', we construct a multi-well energy landscape that generates volume-preserving microstructures during phase change. We use the energy landscapes in (b) and (c) to compute the nucleation and growth of special microstructures during intercalation-induced phase transformation.}
    \label{Fig7}
\end{figure}

\vspace{2mm}
\noindent Fig.~\ref{Fig7}(a) shows combinations of the lattice geometries that satisfy the compatibility criteria for stress-free interfaces and volume-preserving microstructures. For materials undergoing a cubic-to-tetragonal transformation, two independent stretch parameters describe the lattice geometry changes: $\alpha = a/a_0$ and $\beta = c/a_0$. The hatched region in Fig.~\ref{Fig7}(a) identifies combinations of lattice geometries that can form the characteristic twinned domains (i.e., austenite/martensite microstructure) during phase change. The lattice geometries of Li$_{2x}$Mn$_2$O$_4$ (LMO) satisfy this constraint. To achieve stress-free interfaces, the middle eigenvalue of the stretch tensor must satisfy $\lambda_2 = 1$. This condition can be met in two ways: either $\alpha = 1$ for any value of $\beta$, or $\beta = 1$ for any value of $\alpha$. This is shown as vertical ($\alpha = 1$) and horizontal ($\beta = 1$) lines, respectively, in Fig.~\ref{Fig7}(a) and identified by a representative example `A'. A volume-preserving deformation requires the stretch tensor to satisfy $\mathrm{det}\mathbf{U}=1$, which implies $\alpha^2\beta = 1$. This stretch tensor should also satisfy the inequalities to form cubic/twinned-tetragonal interfaces (identified by the hatched region) \cite{ball1987fine,bhattacharya2003microstructure}. The solid line inside the hatched region, therefore, represents combinations of lattice geometries that satisfy the volume-preserving deformation with a representative example identified by `B' in Fig.~\ref{Fig7}(a). These compatibility conditions offer multiple solutions to design lattice geometries of intercalation compounds when compared to the `zero-strain' approach that requires $\alpha=\beta = 1$ as the only solution.\footnote{Cofactor conditions are another set of compatibility constraints that are necessary for improved material reversibility \cite{chen2013study}. For cubic-to-tetragonal transformation, these strong compatibility conditions are only valid for volume fractions $f$ = 0 or $f$ = 1 (equivalent to the $\lambda_2=1$ condition). However, for other symmetry-lowering transformations (e.g., cubic-to-monoclinic) these cofactor conditions can serve as additional design parameters, which when satisfied generates twin microstructures at any arbitrary volume fraction contributing to enhanced reversibility \cite{song2013enhanced}.} 

\vspace{2mm}
\noindent Using two representative lattice geometries `A' and `B' in Fig.~\ref{Fig7}(a)---which satisfy the $\lambda_2=1$ and $|\mathrm{det}\mathbf{U}-1|=0$ conditions respectively---we derive the multi-well energy landscapes for our continuum calculations, see Fig.~\ref{Fig7}(b-c). These energy landscapes have three minima located at $(\bar c,~e_2) = (0.5,~0)$, corresponding to the Li-poor phase, and at $(\bar c,~e_2) = (1.0,~\pm 0.065)$ or $(\bar c,~e_2) = (1.0,~\pm 0.1)$, corresponding to individual lattice variants in the Li-rich phase. The thermodynamic energy barriers governing the phase change in Figs.~\ref{Fig7}(b-c) are constant and correspond to that of Li$_{2x}$Mn$_2$O$_4$ ($0.5 \le x \le 1$). We calibrate the reference phase at $(\bar c,~e_2)=(0.5,~0)$ with the cubic lattice geometry of LiMn$_2$O$_4$ and derive the position of energy wells in Fig.~\ref{Fig7}(b-c) using the lattice deformation gradients of `A' and `B' (see Appendix~C.3 for further details). Note that by tuning the lattice geometries of the transformed phase, we change the separation between energy-wells; however, the thermodynamic energy barriers governing $\mathbf{I}\leftrightarrow\mathbf{U}_i$ transformation are not affected. In the next section, we will show that the position of these energy wells plays a crucial role in generating energy minimizers (i.e., twinned domains or compatible microstructures) that collectively lower the elastic energy barrier during phase change. The elastic energy barrier is not explicitly defined in the continuum energy landscape; however, it emerges during phase transformation (e.g., lattice misfit strains at phase boundaries, volume changes in constrained electrodes). This elastic energy barrier is associated with the dissipation of useful electrochemical work and contributes to the structural decay of electrodes; this energy barrier can be minimized for specific combinations of lattice geometries.
\begin{figure}[ht!]
    \centering
    \includegraphics[width=\textwidth]{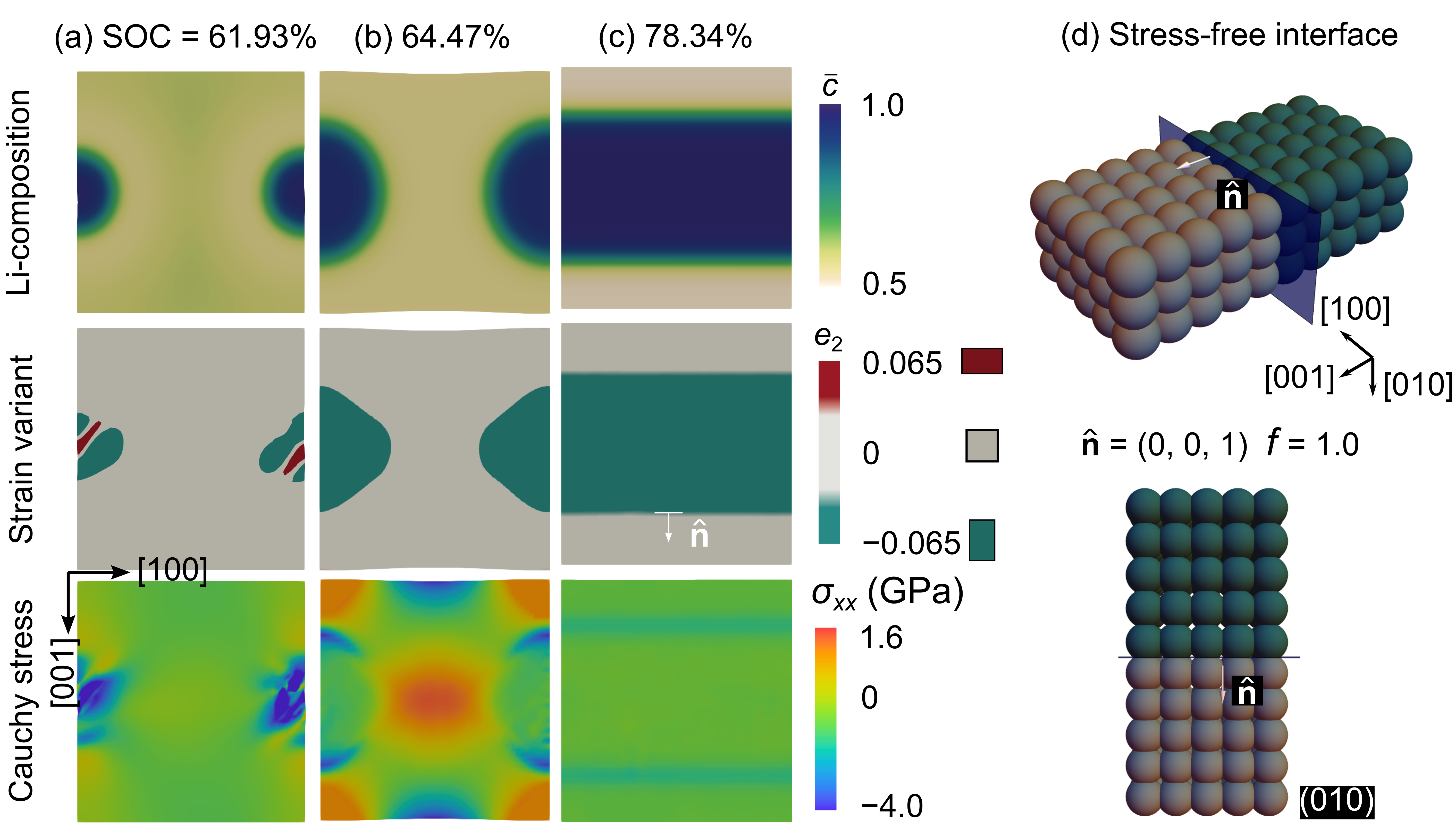}
    \caption{A microstructural evolution pathway for stress-free interfaces, predicted by our multi-variant continuum model. The images on the top, middle and bottom rows, respectively, illustrate the evolution of Li-composition $\bar c$, strain variant $e_2$, and the Cauchy stress component $\sigma_{xx}$, as a function of the SOC. The phase boundary orientation $\hat{\mathbf{n}}$ of the single variant matches the analytical prediction for a stress-free interface in subfigure (d). The peak stress reaches $\sigma_{xx}=1.6$ GPa due to volume changes during nucleation, while the two-phase microstructure generates only negligible stresses of $\sigma_{xx}=-0.3$ GPa across the phase boundary (see subfigure (c)).}
    \label{Fig8}
\end{figure}

\vspace{2mm}
\noindent By minimizing the total energy of the system in the stress-free energy landscape (Fig.~\ref{Fig7}(b)) and under constant flux boundary conditions, stress-free interfaces form in a 500 nm electrode particle. Fig.~\ref{Fig8} shows the sequence of microstructural evolution in an electrode particle satisfying the $\lambda_2 = 1$ constraint (i.e., point `A' in Fig.~\ref{Fig7}(a)). On Li-intercalation into the electrode, a Li-rich phase nucleates at SOC = $61.94\%$, and with continued lithiation, grows through the electrode, forming a planar phase boundary. At the initial nucleation stage, both the tetragonal variants with $e_2 = \pm 0.065$ appear in the lithiated phase, see Fig.~\ref{Fig7}(middle row). This twinned microstructure is transient and a single lattice variant with $e_2 = -0.065$ stabilizes and grows to form a planar interface at SOC = $\sim 77\%$.\footnote{The second type of tetragonal variant $e_2=0.065$ is stabilized and grows to form a compatible interface with the reference phase, when the periodic boundary conditions are applied on the top and bottom edges of the computational domain; this is equivalent to rotating the computational domain by 90 degrees, see Fig.~S2.}

\vspace{2mm}
\noindent The phase boundary between the reference cubic and transformed tetragonal phase has a normal along $\hat{\mathbf{n}}  = (0, \ 0, \ 1)$, see Fig.~\ref{Fig8}(c). This interface orientation predicted by our continuum calculations is consistent with the analytical solution to the compatibility condition in Eq.~(\ref{eq: exact interface}) with $\hat{\mathbf{n}} = (0, \ 0, \ 1)$. The analytical construction of the 3D microstructure in Fig.~\ref{Fig8}(d) shows a planar interface between the cubic-reference phase and a single variant of the tetragonal-transformed phase. This interface is exactly compatible and stress-free according to the sharp interface theory, however our numerical calculations (based on diffuse interface theory) show small interfacial stress of $\sigma_{xx} = -0.3$ GPa, see Fig.~\ref{Fig8}(bottom row). We note that these finite interfacial stresses arise in our continuum calculations because of the non-zero gradient energy terms in Eq.~(\ref{eq:free energy}). These interfacial stresses, however, are negligible when compared to those observed at the LiMn$_2$O$_4$/Li$_2$Mn$_2$O$_4$ phase boundary, see Fig.~\ref{Fig10}(b) for comparison. The compatible and relatively unstressed phase boundary in the crystallographically designed intercalation compound offers the potential to minimize internal stresses during phase transformation and potentially mitigate microcracking \cite{radin2017role}. These stress-free interfaces also lower the energy barriers for reversible transformation, and we discuss this in detail in the Energy Barrier Analysis section.

\begin{figure}[t]
    \centering
    \includegraphics[width=\textwidth]{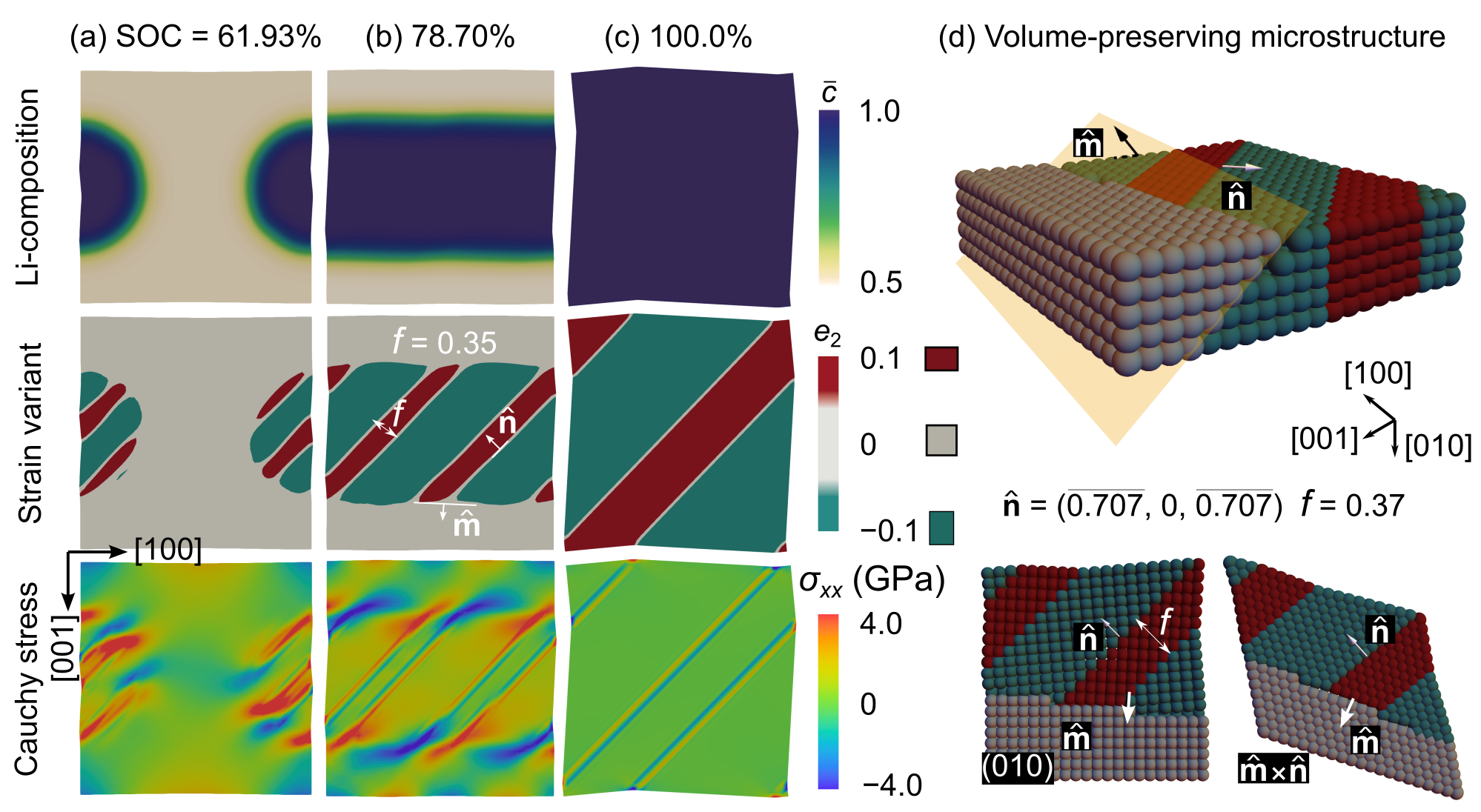}
    \caption{Nucleation and growth of volume-preserving microstructures during a discharge half cycle. (a-c) The top and middle rows show the evolution of Li-composition $\bar c$ and strain variant $e_2$ with increasing SOC. Twinned microstructural features, including the twin plane orientation $\hat{\mathbf{n}}$, volume fraction $f=0.35$, and phase boundary orientation $\hat{\mathbf{m}}$, are highlighted. The bottom row shows the Cauchy stress distribution $\sigma_{xx}$ with tensile stresses reaching up to $\pm4~$GPa across the phase boundary. (d) Analytical construction of 3D microstructures satisfying the volume-preserving ($|\mathrm{det}\mathbf{U}-1|=0$) deformation. The 2D projections of these microstructures onto the (010) and $\hat{\mathbf{m}} \times \hat{\mathbf{n}}$ planes are shown. We note that the geometric features from our continuum simulations deviate by no more than $\pm5.4\%$ from the crystallographic theory.}
    \label{Fig9}
\end{figure}

\vspace{2mm}
\noindent In contrast, intercalation electrodes with lattice geometries satisfying the condition $|\mathrm{det}\mathbf{U}-1|=0$ form twinned microstructures during phase transformation. These microstructures form in materials with symmetry-breaking lattice deformations but can self-accommodate within an electrode, resulting in minimal volume changes. Fig.~\ref{Fig9} shows the nucleation and growth of volume-preserving microstructures in a representative intercalation compound with lattice geometries corresponding to point `B' in Fig.~\ref{Fig7}(a). These microstructures form as a consequence of minimizing the elastic energy across the volume-preserving energy landscape in Fig.~\ref{Fig7}(c). A finely twinned Li-rich phase nucleates at SOC = 61.93$\%$, and with continued lithiation, the Li-rich nucleus grows into a planar microstructure with a twinned volume fraction of $f = 0.35$. 

\vspace{2mm}
\noindent Individual tetragonal lattices satisfy the volume-preserving deformation $|\mathrm{det}\mathbf{U}_i-1|=0$ for $i = 1$ and 3 in Fig.~\ref{Fig9}(a-c); however, the transformed phase is a finely twinned mixture comprising both the variants $\mathbf{U}_1$ and $\mathbf{U}_3$. Although individual lattice variants generate zero-volume changes, the lattice misfit between the reference and transformed phases is significant, which contributes to interfacial stresses at the phase boundary. Minimizing this elastic energy across the multi-well energy landscape in Fig.~\ref{Fig7}(c) generates the finely twinned domains. We note that these microstructures are three-dimensional, and their (010) planar projection introduces errors in the orientation of the phase boundary $\hat{\mathbf{m}}$ and the volume fraction $f$ of the twins, see Fig.~\ref{Fig9}. With these reservations in mind, the geometric features of the microstructures predicted by our continuum simulations, such as the volume fraction $f$, are within $5.4\%$ error with the crystallographic theory. 

\vspace{2mm}
\noindent Fig.~\ref{Fig9}(bottom row) shows the Cauchy stress in the electrode during the nucleation and growth of volume-preserving microstructures. The interfacial stresses are considerable in the volume-preserving microstructure compared to the microstructures satisfying the $\lambda_2 = 1$ condition; however, the hydrostatic stresses $T_\mathrm{H}$ (mean of normal stresses) in Fig.~\ref{Fig10}(a) that correspond to the internal pressure arising from the net volume changes are a minimum in the volume-preserving microstructures compared to both the Li$_{2x}$Mn$_2$O$_4$ and $\lambda_2=1$ interface. These hydrostatic stresses characterize the collective volume expansion of the electrode during phase change, which is minimized by forming volume-preserving microstructures.\footnote{Note, in our 2D computations these hydrostatic stresses are minimized in microstructures with zero area change in the plane of the computational domain, see dashed line in Fig.~\ref{Fig10}(a).} These microstructures offer the potential to mitigate the challenges associated with electrode cracking and delamination failures in solid-state batteries.
\begin{figure}[t]
    \centering
    \includegraphics[width=0.9\textwidth]{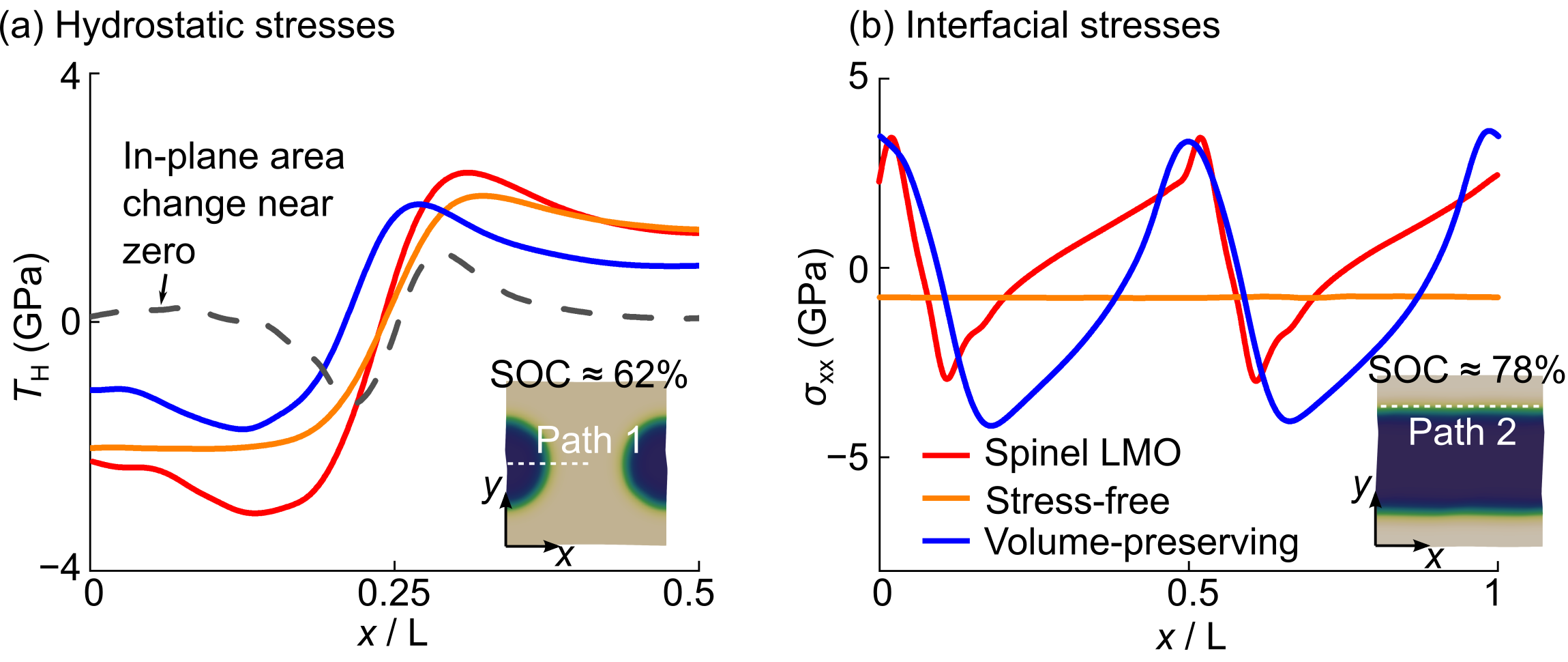}
    \caption{Stress distributions in Li$_{2x}$Mn$_2$O$_4$, stress-free, and volume-preserving microstructures at selected SOCs. (a) We compute the hydrostatic stress distribution $T_\mathrm{H}$, across a Li-rich nucleus at $\mathrm{SOC}\approx 62\%$. Along Path 1, $T_\mathrm{H}$ is small in the volume-preserving microstructure ($\approx 2~\mathrm{GPa}$) compared to the spinel LMO and the stress-free interface. The dashed line represents a hypothetical case with zero area change on the (0, 1, 0) plane, and is associated with much smaller hydrostatic stresses of $T_\mathrm{H}=0.9$ GPa. (b) The Cauchy stress, $\sigma_{xx}$, along the planar phase boundary (Path 2) at $\mathrm{SOC}\approx 78\%$. Significant lattice misfits at phase boundaries in Li$_{2x}$Mn$_2$O$_4$ and volume-preserving microstructures generate stresses of up to $\sigma_{xx} \approx \pm 4$ GPa. In contrast, the stress-free interface, which satisfies the compatibility condition, has minimum stresses of approximately $-0.3$ GPa.}
    \label{Fig10}
\end{figure}

\vspace{2mm}
\noindent Fig.~\ref{Fig10}(b) shows the Cauchy stress distribution $\sigma_{xx}$ along the phase boundary (Path 2). These planar interfaces separate the twinned tetragonal variants of the lithiated phase from the uniform cubic lattices of the reference phase (e.g., in spinel LMO and volume-preserving microstructures). Although these twinned domains are formed to reduce the elastic energy stored at the phase boundary, there is a finite elastic stress along these interfaces as shown in Fig.~\ref{Fig10}(b). The alternating stresses of $\pm 4$ GPa correspond to the misfit between the cubic lattice and the tetragonal lattices. In contrast, the stress-free interface forms an exactly compatible phase boundary between the cubic phase and a single tetragonal lattice variant and thus generates significantly lower stresses of $\sim -0.3$ GPa.

\subsection*{Energy Barrier Analysis}
\label{sec:Energy Barrier Analysis}
\noindent The stresses shown in Figs.~\ref{Fig5}, \ref{Fig8} and \ref{Fig9} are primarily concentrated at the interfaces or phase boundaries. These stresses contribute to the finite elastic energy stored in the system and are a function of both the lattice compatibility (i.e., the fitting together of two phases) and the \textit{emergent} microstructural patterns that form during phase change. These elastic energy contributions at the continuum scale are not captured in the energy barrier of the free energy function (which are calibrated with the material's thermodynamic and elastic constants) but instead emerge during phase transformation. In this section, we show that this elastic energy contribution can be reduced by energy-minimizing sequences, such as twins and stress-free interfaces ($\lambda_2 = 1$), and doing so has a direct impact on the chemical potential (related to voltage hysteresis) and driving forces (related to material reversibility) in intercalation compounds. In these analyses, we do not assume a pre-existing nucleus of the transformed phases, but instead compute the energy barriers and analyze the stability of a two-phase microstructural pattern that phenomenologically forms during charge/discharge processes. 
\begin{figure}[ht!]
    \centering
    \includegraphics[width=\textwidth]{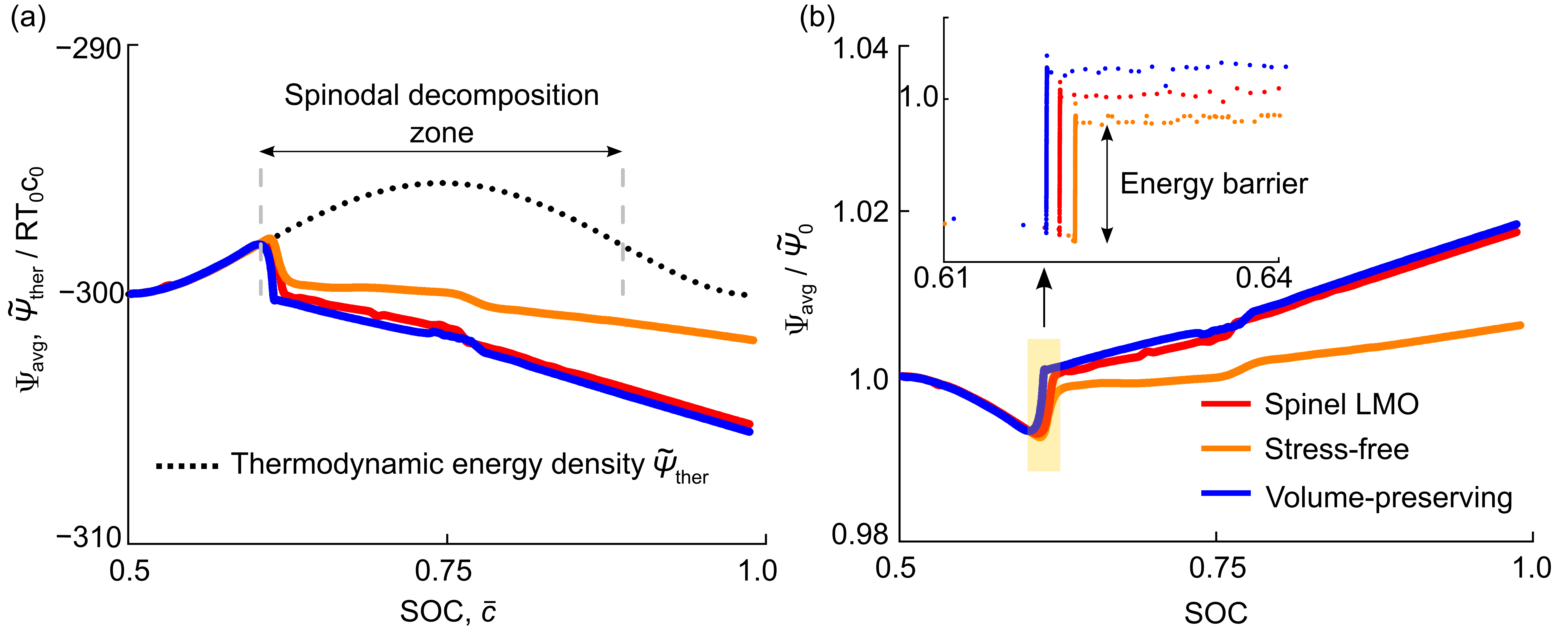}
    \caption{(a) We compute the total free energy of the system $\Psi_{\mathrm{avg}}$ as a function of the SOC. This energy $\Psi_{\mathrm{avg}}$ increases up to the first spinodal point, after which the system phase separates and is marked by an abrupt drop in the total energy. This drop in energy is different for each of the three cases, namely Li$_2$Mn$_2$O$_4$, stress-free interface, and volume-preserving microstructure, and is a consequence of the collective elastic energy barriers that emerge during phase change. (b) Normalized energy barriers during phase change in the three microstructures ($\tilde{\psi}_0$ is the average energy of the Li-poor phase). The inset plot shows that the stress-free interface offers the lowest nucleation energy barrier during phase change (because of the exactly compatible phase boundaries) when compared to the energy barriers that emerge in Li$_2$Mn$_2$O$_4$ and volume-preserving $|\mathrm{det}\mathbf{U}-1|=0$ microstructures.}
    \label{Fig11}
\end{figure}

\vspace{2mm}
\noindent Fig.~\ref{Fig11} shows the free energy landscape governing phase change in three representative materials, namely Li$_{1-2}$Mn$_2$O$_4$, and two crystallographically designed compounds satisfying the stress-free ($\lambda_2 = 1$) interface and volume-preserving ($|\mathrm{det}\mathbf{U}-1|=0$) deformation, respectively. To compare energy barriers across the three representative materials, we construct the energy landscape using identical thermodynamic and elastic constants (corresponding to those of LiMn$_2$O$_4$). For example, in Fig.~\ref{Fig11}(a) the dashed line corresponds to the thermodynamic energy barrier governing the phase change ($\bar{c} = 0.5\to1.0$) is calibrated with the open-circuit voltage of Li$_{1\to2}$Mn$_2$O$_4$ \cite{thackeray1983lithium} and is held constant across the three cases. Likewise, the elastic constants including the bulk modulus ($\mathrm{K}=\frac{c_{11}+c_{12}}{2}$), deviatoric modulus ($\mathrm{C}=\frac{c_{11}-c_{12}}{2}$) and shear modulus ($\mathrm{G}=c_{44}$) are fitted with those of LiMn$_2$O$_4$. This enables us to compare the differences in elastic energy barriers that emerge during phase change across the three representative materials.

\vspace{2mm}
\noindent Fig.~\ref{Fig11}(a) shows the total energy ($\Psi_{\mathrm{avg}}=\int_\Omega(\tilde\psi_{\mathrm{ther}}+\psi_{\mathrm{elas}}+\psi_{\mathrm{grad}})dV/V$) of the system for the three cases as a function of the intercalant composition $\bar{c}$. In all cases, the total energy increases until the spinodal point, beyond which the domain decomposes into a two-phase microstructure separated by a phase boundary. The lattice compatibility across these phase boundaries differs among the three cases (governed by the deformation gradient) and manifests as distinct interfacial stresses in our calculations (see Cauchy stress plots in Figs.~\ref{Fig5}, \ref{Fig8} and \ref{Fig9}). These stresses contribute to the finite elastic energy in the respective materials, which, in turn, generates additional energy barriers that hinder phase separation. 

\vspace{2mm}
\noindent The collective energy barriers for each of the three representative materials are shown in Fig.~\ref{Fig11}(b). The positive slope of these curves corresponds to the finite elastic energy accumulated in the system during the phase change. This elastic energy in Li$_2$Mn$_2$O$_4$ and the volume-preserving microstructure corresponds to the interfacial stresses at the phase boundary. The twinned domains relieve the lattice misfit at the phase boundary and lower the interfacial stresses; however, the resulting phase boundary is still not exactly compatible. These small but finite stresses contribute to elastic energy in the system that manifests in the form of a positive slope in the energy barrier curves for Li$_2$Mn$_2$O$_4$ and the volume-preserving microstructure, see Fig.~\ref{Fig11}(b). In contrast, the elastic energy barrier is the smallest for the stress-free ($\lambda_2=1$) interface. In this case, a single variant of the transformed phase forms an exactly compatible interface with the reference phase. This interface is associated with theoretically zero stresses, which lowers the elastic energy stored in the system.\footnote{The $\lambda_2=1$ interface is an exactly compatible interface, however, in our diffuse-interface modeling we note finite stresses ($-0.3$ GPa) during phase change, see Fig.~\ref{Fig8}. These stresses correspond to the energy penalty for gradients of the composition order parameter $\nabla\bar{c}$ and the strain order parameter $\nabla e_2$ in our continuum framework.} Therefore, given that all other thermodynamic and elastic constants are identical in the three representative materials, the elastic energy barriers emerging during phase change are the lowest for the $\lambda_2=1$ interface.\footnote{We can further minimize this elastic energy barrier by designing lattice deformations that not only satisfy the $\lambda_2=1$ criterion, but also the volume-preserving ($|\mathrm{det}\mathbf{U}-1|=0$) deformation \cite{bhattacharya2004crystal}. For the case of cubic-to-tetragonal transformation, this would reduce to a rigid lattice constraint with zero distortion.}

\begin{figure}[ht!]
    \centering
    \includegraphics[width=\textwidth]{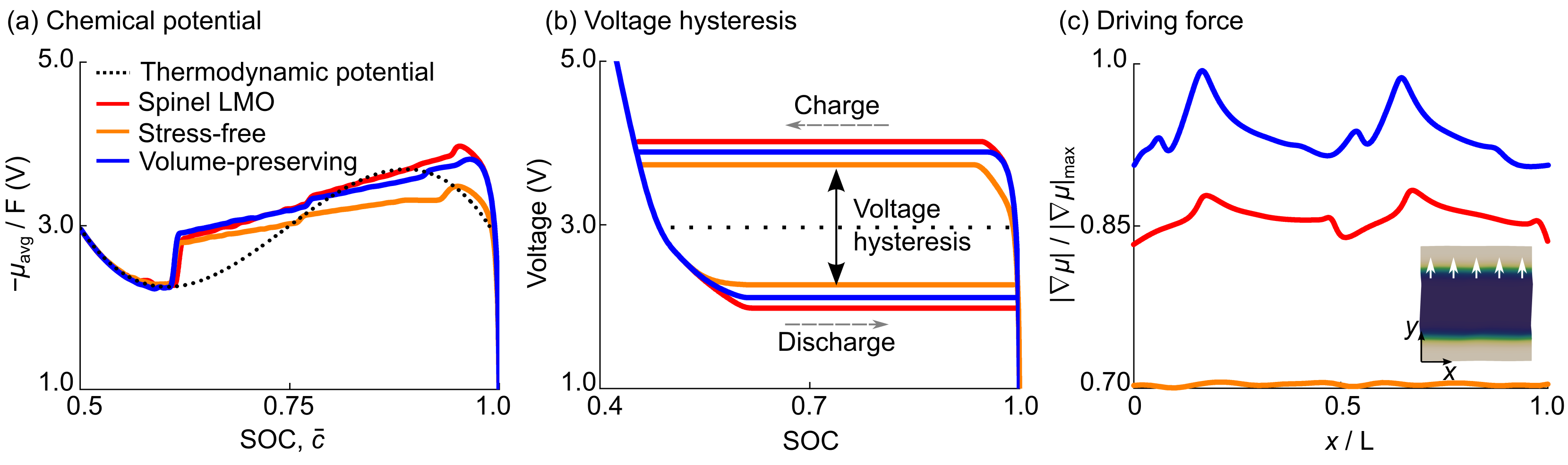}
    \caption{(a) Chemical potential curves from our continuum calculations for the three representative materials (namely Li$_{1-2}$Mn$_2$O$_4$ and two crystallographically designed compounds satisfying the stress-free $\lambda_2 = 1$ interface and volume-preserving $|\mathrm{det}\mathbf{U}-1|=0$ deformation). All curves show a positive slope, indicating a finite elastic energy stored at the phase boundaries. This slope is the smallest in the $\lambda_2 = 1$ interface because of near-zero interfacial stresses, but is noticeably larger in the other two materials. (b) We derive voltage hysteresis loops from our continuum calculations for the three representative materials. The width of the voltage hysteresis loop is the smallest for the $\lambda_2=1$ interface, and shows relatively wider loops for Li$_2$Mn$_2$O$_4$ and the $|\mathrm{det}\mathbf{U}-1|=0$ deformation. We attribute the narrow hysteresis loop in the $\lambda_2 =1 $ interface to the smaller elastic energy stored at the phase boundaries. These results offer preliminary insights into the advantages of crystallographically designing intercalation compounds to lower energy barriers that govern phase transformation. (c) We calculate the driving forces across the phase boundary during phase change. These forces are 30$\%$ lower for the $\lambda_2=1$ interface when compared to Li$_2$Mn$_2$O$_4$. Smaller driving forces reduce energy dissipation and promote structurally reversible phase transformations.}
    \label{Fig12}
\end{figure}

\vspace{2mm}
\noindent These energy barriers accompanying phase change have a direct impact on the chemo-mechanical performance of intercalation compounds. For example, the elastic energy barriers alter the chemical potential values (governing the two-phase separation) and thus affect the width of the voltage hysteresis loops, see Fig.~\ref{Fig12}(a-b). For reference, the dashed line Fig.~\ref{Fig12}(a) shows the thermodynamic potential of Li$_{2x}$Mn$_2$O$_4$ (0.5 $\leq x \leq$ 1), and the dashed line in Fig.~\ref{Fig12}(b) corresponds to the open-circuit voltage measurement \cite{thackeray1983lithium}. The chemical potential curves for the three cases deviate with a positive slope from the voltage plateau in Fig.~\ref{Fig12}(a). This deviation is significant in Li$_2$Mn$_2$O$_4$ and the volume-preserving microstructure, when compared to the $\lambda_2=1$ interface. We attribute this difference in the slopes to the larger elastic energy stored at the phase boundaries in Li$_2$Mn$_2$O$_4$ and volume-preserving material systems. The near-zero interfacial stresses in the $\lambda_2=1$ interface generate a relatively small elastic energy that governs the phase change. This behavior mirrors that of other intercalation compounds (e.g., LiFePO$_4$ \cite{cogswell2012coherency} and LiCoO$_2$ \cite{nadkarni2019modeling}), which form solid solutions at high C-rate, resulting in flatter curves as the coherency strain becomes less influential. Our key finding from Fig.~\ref{Fig12} is that fine-tuning the lattice compatibility of individual lattices to achieve the stress-free interface during phase transformation---in addition to the thermodynamic and kinetic driving forces---lowers the energy barrier and internal stress, thereby improving the performance of intercalation compounds.

\vspace{2mm}
\noindent Fig.~\ref{Fig12}(b) shows the voltage hysteresis loops for the three representative materials. In deriving these loops, we assume symmetric energy barriers that govern the charge/discharge processes. The area enclosed by the voltage hysteresis loops quantifies the energy lost during an electrochemical cycle \cite{van2022hysteresis}. Among the three cases, the voltage hysteresis loop is the narrowest for the $\lambda_2=1$ interface. This suggests that the phase change in the $\lambda_2=1$ interface is driven by relatively small differences in voltage values and is associated with a minimum energy dissipation. In contrast, the voltage hysteresis loops for Li$_2$Mn$_2$O$_4$ and the volume-preserving microstructure are wider, indicating that larger voltage drops are necessary to induce a phase change. 

\vspace{2mm}
\noindent The voltage hysteresis calculation for Li$_2$Mn$_2$O$_4$ aligns with the $2-4$~V voltage window reported in the experiments by Erichsen et al. \cite{erichsen2020tracking}. The width of this voltage hysteresis loop can be further reduced by minimizing the thermodynamic energy barrier associated with $\psi_{\mathrm{ther}}$ in Eqs.~(\ref{eq:3D_free_energy}) and (\ref{eq:free energy}), see Appendix~C.4. This thermodynamic energy barrier $\Delta \psi$ is a function of temperature (that is, $\Delta \psi \propto -T$) and the enthalpy of mixing coefficients $\alpha_i$. Increasing the temperature, or equivalently decreasing $\alpha_1$ reduces the barrier height and contributes to narrower voltage hysteresis loops; see Fig.~S1. This is analogous to the regular solution model, where reducing the enthalpy of the mixing parameter $\Omega$ or increasing the temperature $T$ proportionally lowers the thermodynamic energy barrier. We detail this analysis in Appendix~C.4.

\vspace{2mm}
\noindent Fig.~\ref{Fig12}(c) shows the maximum driving forces acting across the phase boundaries in the three representative materials at SOC $\approx 78\%$. The total driving force for phase change $\nabla \mu$  is calculated from Eq.~(\ref{eq:total chemical potential}), in which the total free energy of the system, including the elastic energy effects in Eq.~(\ref{eq:elastic driving force}), is accounted for. A small driving force demonstrates that the phase boundary moves back and forth in the material for the smallest perturbation (or energy changes) during charge/discharge processes. At SOC $\approx 78\%$, the morphology of the phase boundary in all three cases is a planar interface (see inset Fig.~\ref{Fig12}(c) showing the direction of the driving force). The magnitude of the driving force acting across the phase boundary is highest in the volume-preserving microstructure (with $|\mathrm{det}\mathbf{U}-1| = 0$) and lowest in the stress-free interface (with $\lambda_2 = 1$). The driving force in Li$_2$Mn$_2$O$_4$ (with $|\mathrm{det}\mathbf{U}-1|\approx 0.064$ and satisfying the average compatibility condition in Eq.~(\ref{eq:AM interface})) is intermediate between the three cases. Their magnitude corresponds to the energy required to drive the phase transformation. In other words, the lower driving force in the $\lambda_2=1$ case suggests that phase boundaries can be moved with minimal energy input, leading to reduced energy dissipation and enhanced reversibility during phase change.

\section*{Discussion}
\noindent Our multi-variant continuum model extends existing phase-field methodologies by predicting crystallographic microstructures that form during phase transformations in intercalation compounds. A distinguishing feature of our approach is that we construct a multi-well energy landscape that is a function of not only the Li-composition but also individual lattice variants of a given phase. By minimizing the total free energy (for Li$_2$Mn$_2$O$_4$), we predict geometrically accurate microstructural patterns, such as twin plane orientations, volume fractions, and compatible phase boundaries that emerge during phase transformation. These energy-minimizing deformations were not previously captured by phase-field models for intercalation compounds, and offer new routes to design materials with improved structural reversibility.

\vspace{2mm}
\noindent Our key finding is that by designing lattice compatibility in intercalation compounds, we can significantly reduce the energy barriers governing phase transformations. These barriers manifest primarily as elastic energy stored as interfacial stresses in microstructural patterns, which can be minimized in materials with special combinations of lattice geometries (e.g., $\lambda_2=1$ microstructure reduces interfacial stresses from $\pm4\mathrm{~GPa}$ to $-0.3\mathrm{~GPa}$, and the $|\mathrm{det}\mathbf{U}-1|=0$ microstructure reduces hydrostatic stresses to near-zero values). These crystallographically designed materials have small energy barriers; therefore, they require substantially lower driving forces (by 30$\%$)  to move phase boundaries and are associated with narrower voltage hysteresis loops. These elastic energy barriers accompanying phase transformations are difficult to estimate using atomistic or first-principles calculations, but are effectively captured in our continuum approach.

\vspace{2mm}
\noindent Additionally, we show that the twin boundaries that form during phase change serve as conduits for fast Li-ion transport. The finite stresses along the twin interfaces and the phase boundaries act as additional driving forces for Li-diffusion. These driving forces are anisotropic, facilitating faster diffusion along the twin interfaces when compared to the diffusion across the twins. This finding highlights another potential of crystallographically designing intercalation compounds in which twin interfaces are engineered radially outward in electrode particles to facilitate rapid charging/discharging. 

\vspace{2mm}
\noindent Our ability to predict accurate elastic energy barriers and their effects on macroscopic material properties is limited. This limitation arises from the lack of experimental characterization of interfacial energy constants and the strain energy landscape, as well as the computational costs associated with 3D calculations. These factors are likely to affect the elastic transition layer (i.e., the phase boundary) predicted by our model. With ongoing efforts to characterize interfacial constants \cite{lim2016origin,temprano2024advanced} and the use of first-principles calculations to derive free-energy landscapes \cite{puchala2023casm}, we expect to gain quantitative insights into these coefficients and improve the accuracy of our predictions. Additionally, by optimizing our numerical algorithms and/or porting to GPUs to efficiently solve higher-order PDEs, our continuum model will enable us to compute larger domain sizes and to investigate the interplay between lattice compatibility and nonlinearities (e.g., defects such as grain boundaries and dislocations) during phase change.

\vspace{2mm}
\noindent We formulate our multi-variant continuum model using the Cauchy-Born rule, which relates the movement of atoms to the overall deformation of a solid. This rule works well for simple crystals with Bravais lattices but may fail for compounds with multi-lattices (e.g., perovskite Li$_x$ReO$_3$ \cite{bashian2018correlated}). Unit cells of intercalation compounds, such as Li$_2$Mn$_2$O$_4$, NaMnO$_2$, and Li[Li$_{1/3}$Ti$_{5/3}$]O$_4$ \cite{erichsen2020tracking,abakumov2014multiple, ohzuku1995zero}, can be described as Bravais lattices, and our continuum model accurately predicts microstructural patterns that are consistent with experimental observations. However, applying this framework to intercalation compounds with multi-lattice unit cells (i.e., containing two or more constituent Bravais lattices such as the perovskite Li$_x$ReO$_3$ \cite{bashian2018correlated}) requires careful consideration. For example, Li insertion into perovskite ReO$_3$ induces internal rotations of the individual Re octahedra, making lattice mapping based solely on the deformation gradient insufficient. In these multi-lattices, atoms may shift and/or shuffle during displacive phase transformations, and these movements may not be fully captured by the deformation gradient alone. Under dynamic or high-rate loading conditions, atomic shuffling may affect material response (e.g., hysteresis, reversibility, or diffusion). In such cases, we need to investigate whether the relative shifts between Bravais lattices are congruent and whether the microstructural evolution can be modeled under quasi-equilibrium conditions. 

\vspace{2mm}
\noindent Despite these challenges, our computations remain valuable for understanding symmetry-breaking phase transformations at the continuum scale. The model quantitatively captures the interplay between individual lattice deformations and Li-flux during phase change. By constructing a multi-well energy landscape, we allow for individual lattices to deform into symmetry-related variants and predict energy-minimizing sequences (e.g., twins, compatible phase boundary). In intercalation compounds, these microstructures evolve under dynamic loading conditions, making them a nonlinear problem that is difficult to solve analytically. Our multi-variant continuum model not only investigates this nonlinear behavior, but also predicts quantitative elastic energy barriers that manifest during phase change (e.g., decrease in interfacial stresses for compatible microstructures in comparison to Li$_2$Mn$_2$O$_4$) and their impact on driving forces and hysteresis in an electrochemical environment. These predictions of crystallographic microstructures are phenomenological, made without a priori assumptions about the critical Li-rich nucleus or explicit boundary conditions on moving interfaces. The equilibrated microstructures from our continuum model are geometrically accurate and are remarkably close (within $\pm5\%$ error) to experimental observations of crystallographic microstructures in intercalation compounds. This ability to make quantitative predictions under dynamic electrochemical loads enables us to use our continuum model as both an exploratory tool to investigate fundamental microstructural mechanisms in symmetry-breaking phase transformations and as a design tool to crystallographically engineer intercalation compounds with improved reversibility.

\vspace{2mm}
\noindent The proof of concept of our crystallographically designed materials (e.g., satisfying the conditions $\lambda_2 = 1$, $|\mathrm{det}\mathbf{U}-1|=0$) expands the design parameter space beyond the existing zero-strain design rule for intercalation compounds. Unlike the zero-strain condition $\mathbf{F}=\mathbf{I}$ commonly used in the literature \cite{bonnick2018insights, yang2022new, zhao2022zero,zhao2022design}, our compatibility design rules do not restrict lattices from deforming but instead impose less stringent constraints on lattice deformation to satisfy interfacial and volume compatibility \cite{zhang2023designing, renuka2022crystallographic,zhang2024coupling}. In doing so, we find multiple solutions for lattice geometries, see Fig.~\ref{Fig7}(a), which enable chemists and material scientists to synthesize intercalation compounds (e.g., site-selective doping) over a wider composition space. Furthermore, our multi-variant continuum model can be extended to other symmetry-lowering transformations in intercalation cathodes such as NaMnO$_2$ \cite{abakumov2014multiple}, Prussian blue analogues \cite{wang2020reversible,zhang2022lithiated}, and perovskite solid electrolytes (Li$_{3x}$La$_{2/3-x}$TiO$_3$) \cite{lu2021perovskite}. To describe other symmetry-lowering transformations (e.g., orthorhombic to monoclinic in NaMnO$_2$ \cite{abakumov2014multiple}), our model can be further generalized by incorporating shear strain order parameters (e.g., e$_4$, e$_5$, and e$_6$) in addition to e$_2$ and e$_3$. Our computations predict the length scales and electrochemical boundary conditions under which the crystallographic microstructures nucleate and grow steadily in these materials. These continuum findings provide insights into identifying potential electrode particle sizes and morphologies. Our findings on the impact of crystallographic microstructures on electrochemical performance (e.g., voltage hysteresis loops, driving forces) further demonstrate that lattice compatibility, in addition to the thermodynamic and kinetic barriers, plays an important role in the performance and reversibility of intercalation compounds. 

\vspace{2mm}
\noindent Beyond intercalation compounds, our theoretical and computational framework will serve as an investigatory tool to study the nucleation of martensitic microstructures in shape-memory alloys \cite{tuuma2016size}, understand ferroelastic toughening in ceramics \cite{jetter2019tuning}, and to investigate the deformation-microstructure-property relations in other phase-transformation materials (e.g., ferroelectrics \cite{renuka2016nanoscale}, magnets \cite{guan2025hysteresis, balakrishna2022compatible, renuka2022design}, molecular crystals \cite{tiwari2025micromechanical}). Previous analytical research on these materials has led to remarkable progress in reducing fatigue and improving reversibility \cite{chluba2015ultralow,gu2018phase,zarnetta2010identification,song2013enhanced}, but has relied on assumptions of specific nucleus geometries in equilibrium \cite{knupfer2013nucleation, chen2013study, zhang2009energy}. In our framework, we allow for these crystallographic features to evolve in-situ and without a priori assumptions, and thus establish a framework to quantitatively probe the energy barriers emerging during phase change. These continuum techniques could be powerful in investigating the origins of irreversibility in phase-transformation materials.

\vspace{2mm}
\noindent In summary, we present a materials design strategy centered on lattice compatibility and reducing energy barriers to enhance the lifespan and electrochemical performance of intercalation compounds. Using Li$_2$Mn$_2$O$_4$ as a representative material, we develop a multi-variant continuum model that directly links lattice deformations with the crystallographic microstructures formed during phase change. Our continuum model not only predicts geometrically accurate microstructures in Li$_2$Mn$_2$O$_4$, but also demonstrates the beneficial role of twin interfaces in promoting faster Li-ion transport and minimizing interfacial stresses. Our energy barrier analysis shows that compatible interfaces lower the elastic energy barriers, resulting in smaller driving forces for phase transformation and narrower voltage hysteresis loops. As a proof of concept, we use the model to design intercalation compounds with stress-free interfaces and volume-preserving microstructures, which, in turn, enable easier phase transformations and narrower voltage hysteresis. These results suggest that tailoring lattice deformations to meet special compatibility conditions---rather than suppressing them entirely---can result in compounds that maintain structural integrity and electrochemical performance over extended use.

\section*{Methods}
In this section, we describe the continuum theory of crystalline solids undergoing a first-order phase transformation and highlight its application to intercalation compounds. Consider a crystalline solid $\Omega$ in a three-dimensional space $\mathbb{R}^3$. In the reference configuration, each point in the solid is defined by a position vector $\mathbf{x}$. During phase transformation, a part of this solid deforms according to $\mathbf{y}(\mathbf{x})$ and the deformation gradient is defined by the tensor $\mathbf{F} = \nabla\mathbf{y}$. In the Results section of the main paper, we showed that this overall deformation gradient is related to the displacive movement of atoms via the Cauchy-Born rule \cite{ericksen2008cauchy}, and can be decomposed into a unique rotation and a stretch tensor: $\mathbf{F = QU}$.

\subsection*{Stretch Tensor}
\begin{figure}[t]
    \centering
    \includegraphics[width=0.95\textwidth]{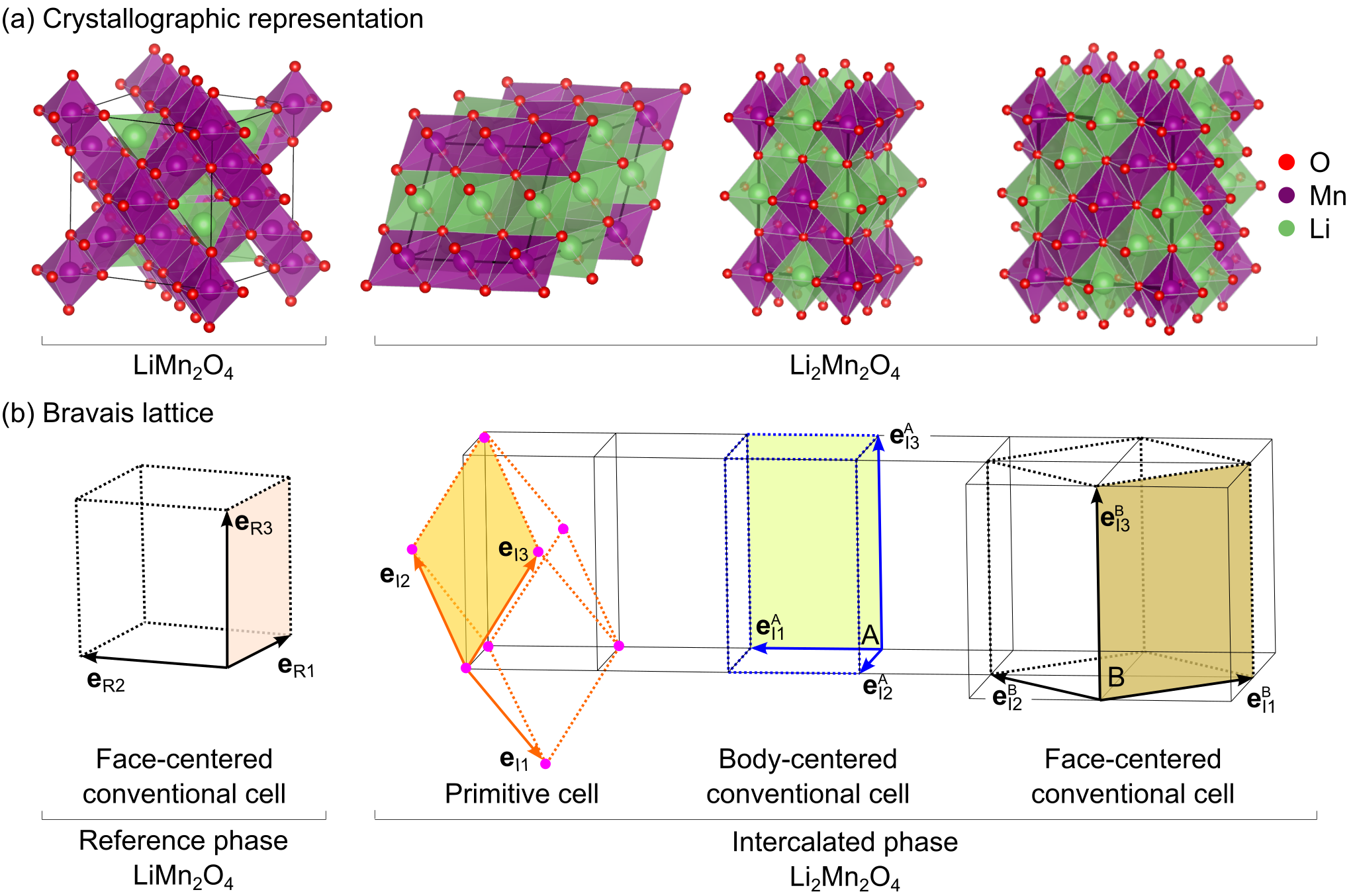}
    \caption{(a) Crystallographic and (b) Bravais lattice representations of unit cells of LiMn$_2$O$_4$ (reference phase) and Li$_{2}$Mn$_2$O$_4$ (intercalated phase). The vectors $\{\mathbf{e}_\mathrm{R1}, \mathbf{e}_\mathrm{R2}, \mathbf{e}_\mathrm{R3}\}$ span the unit cell of LiMn$_2$O$_4$ ($Fd3m$, $a_0$ = 8.24 \text{\AA}). Similarly, vectors $\{\mathbf{e}_{\mathrm{I}1}, \mathbf{e}_{\mathrm{I}2}, \mathbf{e}_{\mathrm{I}3}\}$ enclose the primitive unit cell of the intercalated phase Li$_{2}$Mn$_2$O$_4$. Representative examples of conventional unit cells are identified by vectors with superscripts `A' and `B'. On minimizing the distance function Eq.~(\ref{eq:distance function}) across the three stretch tensors mapping LiMn$_2$O$_4$ to primitive, body-centered, and face-centered Li$_{2}$Mn$_2$O$_4$, respectively, we identify that the stretch tensor corresponding to the face-centered conventional cell of Li$_2$Mn$_2$O$_4$ ($F4_1/ddm$, $a = 8.0$ Å, $c = 9.3$ Å) has the smallest structural distortion. We use this stretch tensor as an input for our model.}
    \label{Fig1}
\end{figure}

\noindent The stretch tensor $\mathbf{U}$ for a given material can have multiple solutions based on the choice of lattice vectors used to describe the Bravais lattices of the reference and transformed phases. For example, consider the structural transformation of the lattices accompanying the LiMn$_2$O$_4$ to Li$_2$Mn$_2$O$_4$ intercalation process. Assume a cubic unit cell for the reference LiMn$_2$O$_4$ phase described by lattice vectors $\mathbf{e}_{\mathrm{R}i}$, see Fig.~\ref{Fig1}. The unit cells of the transformed Li$_2$Mn$_2$O$_4$ can be defined in multiple ways, such as $\mathbf{e}_{\mathrm{I}i}$ for a primitive cell, $\mathbf{e}^{\mathrm{A}}_{\mathrm{I}i}$ or $\mathbf{e}^{\mathrm{B}}_{\mathrm{I}i}$ for conventional cells, see Fig.~\ref{Fig1}. The stretch tensors mapping the reference to the transformed lattices can therefore have multiple solutions. Determining the optimal stretch tensor for a given phase transformation is important for developing our continuum theory.

\vspace{2mm}
\noindent To determine an optimal stretch tensor, we use a structural transformation algorithm developed in recent works \cite{chen2016determination, zhang2023designing}. In this algorithm, a distance function is introduced to quantify the lattice distortion between all combinations of reference and transformed phases:
\begin{align}
    \mathrm{dist}(\mathbf{U})=||\mathbf{U}^{-2}-\mathbf{I}||^2
    \label{eq:distance function}
\end{align}

\noindent The stretch tensor minimizes the distance function in Eq.~(\ref{eq:distance function}) is chosen as the optimal stretch tensor for constructing continuum microstructures. This approach to identifying the stretch tensor has seen recent success in the context of shape-memory alloys \cite{chen2016determination}, and more recently, in our work on intercalation compounds \cite{zhang2023designing}. 

\vspace{2mm}
\noindent By applying this algorithm to Li$_2$Mn$_2$O$_4$ we identify the cubic ($a_0$ = 8.24 \text{\AA}) to tetragonal ($a = 8.0$ \AA, $c = 9.3$ \AA) deformation as the distance-minimizing stretch tensor. This transformation pathway is consistent with experimental results \cite{erichsen2020tracking, luo2020operando}. We follow this approach to determine stretch tensors for other representative intercalation compounds, including NaMnO$_2$ \cite{abakumov2014multiple} and Li[Li$_{1/3}$Ti$_{5/3}$]O$_4$ \cite{ohzuku1995zero}, see Tables~S3 and S4 in the Supplement.

\subsubsection*{Kinematic Compatibility Conditions}
\label{sec:Kinematic Compatibility Conditions}
\noindent The compatibility conditions predict the orientation of the twin plane along which lattices with different deformation gradients (or, equivalently, the stretch tensors) align to form a compatible interface. In Eq.~(\ref{eq:twin equation}) of the main paper, two tetragonal lattices, $\mathbf{U}_I$, $\mathbf{U}_J$, generated during the cubic-to-tetragonal transformation in Li$_2$Mn$_2$O$_4$ rotate and fit together to form a compatible interface (i.e., $\mathbf{QU_\mathit{J}-U_\mathit{I}} = \mathbf{a\otimes\mathbf{\hat{n}}}$). The solutions to the compatibility condition are vectors $\mathbf{a}$, $\hat{\mathbf{n}}$ that describe the orientation of the twin planes. These twin planes are special interfaces along which lattices with distinct deformation gradients meet and share an edge that is relatively unstretched. This compatible fit between neighboring lattices results in coherent twin interfaces and allows several twins to form in a material with no (or minimal) energy penalty. However, twin interfaces are rarely found in isolation; instead, several parallel twins appear as fine mixtures during phase transformation (see Figs.~\ref{Fig0} and \ref{Fig14}(a)).

\vspace{2mm}
\noindent The finely twinned mixtures form an interface with a uniform domain, see Fig.~\ref{Fig14}(a). This interface, commonly referred to as the austenite/martensite interface (terminology used in shape-memory alloys), forms as a consequence of energy minimization and satisfies the compatibility condition in Eq.~(\ref{eq:AM interface}), $\mathbf{Q}'(f\mathbf{QU}_{J}+(1-f)\mathbf{U}_{I}) = \mathbf{I}+\mathbf{b}\otimes\mathbf{\hat{m}}$. The uniform reference phase, with a single crystal structure, is represented by an identity stretch tensor $\mathbf{I}$, and the finely twinned transformed phase consists of parallel twins between the tetragonal variants $\mathbf{U}_I$ and $\mathbf{U}_J$. The interface between any single tetragonal variant and the reference phase is elastically stressed because of the lattice mismatch. To minimize this interfacial energy, the tetragonal lattices form a finely twinned mixture that is nearly compatible with the reference phase. The volume fraction of the twins $0 \leq f \leq 1$ is a sophisticated averaging of the tetragonal variants that minimizes the elastic energy stored at the austenite/martensite interface. The vectors $\mathbf{b,\hat{m}}$ describe the orientation of this interface, and the rotation tensors are given by $\mathbf{Q}',\mathbf{Q}$ \cite{ball1987fine}. 

\vspace{2mm}
\noindent The austenite/martensite microstructure is three dimensional in nature and inherently complex to visualize; specfically, the surface normal of the twin interface in the martensite phase, $\hat{\mathbf{n}}$, does not lie in the same plane as the surface normal of the austenite/martensite interface, $\hat{\mathbf{m}}$. However, there exists a plane in $\mathbb{R}^3$ that is orthogonal to both the twin interface and the austenite/martensite phase boundary---namely the vector $\hat{\mathbf{m}}\times \hat{\mathbf{n}}$---on which 2D projections of the microstructural features can be geometrically represented. We note that the experimental images in Fig.~\ref{Fig14}(b) are projected on the $(0,~1,~0)$ 2D plane \cite{erichsen2020tracking}, which introduces geometric inaccuracies in the crystallographic features, see Figs.~\ref{Fig14}(b) and \ref{Fig5}(e). These errors in twin plane orientations ($\Delta \theta \approx 0.89^\circ$) and volume fractions ($\Delta f \approx 0.01$) are small in the case of Li$_2$Mn$_2$O$_4$, due to its small out-of-plane strain components.

\subsubsection*{Special Microstructures}
Recent advances in energy barrier analysis have identified special combinations of lattice geometries that generate stress-free interfaces and volume-preserving (or self-accommodating) microstructures \cite{knupfer2013nucleation, zhang2009energy, balakrishna2021tool}. These microstructures, when engineered to form in ferroelastic materials (e.g., shape-memory alloys, ferroelectric, ferromagnets, semiconductors) \cite{liang2020tuning,dubey2020improved,srivastava2011direct,balakrishna2022compatible,gu2018phase}, have significantly improved the structural reversibility of these materials during phase change. 

\vspace{2mm}
\noindent In this section, we outline the theory underlying these special microstructures and identify combinations of lattice geometries necessary to form (a) stress-free interfaces (or $\lambda_2 = 1$) and (b) volume-preserving microstructures (or $|\mathrm{det}\mathbf{U}-1|=0$).\footnote{Note, materials undergoing a cubic-to-tetragonal phase transformation with a stretch tensor $\mathbf{U}$ can form volume-preserving (or self-accommodating) microstructures when $|\mathrm{det}\mathbf{U}-1|=0$. Other symmetry-lowering phase transformations require satisfying additional constraints as detailed in Chapter 9 of Ref.~\cite{bhattacharya2003microstructure}.} These parameters will serve as inputs for our multi-variant continuum model and the Energy Barrier Analysis section.

\vspace{2mm}
\noindent A phase boundary is \textit{stress-free} when the lattices undergo a finite deformation but share an edge that remains relatively unstretched between the two phases (i.e., satisfying the $\lambda_2 = 1$ criterion). This condition corresponds to an exact interface (without misfit strains) between the reference and transformed lattices and is associated with enhanced reversibility and low fatigue in phase transformation materials \cite{song2013enhanced, chluba2015ultralow,chen2013study}. The compatibility condition for this exact interface is given by:
\begin{align}
    \mathbf{QU-I=a}\otimes\hat{\mathbf{n}}.
    \label{eq: exact interface}
\end{align}

\noindent The change in lattice geometry between the reference and transformed phases is described by the stretch tensor $\mathbf{U}$ and a rotation tensor $\mathbf{Q}$. The transformed lattices form a compatible interface with the reference phase $\mathbf{I}$ if the eigenvalues of the stretch tensor $\mathbf{U}$ satisfy $\lambda_1 \leq \lambda_2 = 1 \leq \lambda_3$. The middle eigenvalue criterion $\lambda_2=1$ corresponds to the common lattice edge between the reference and transformed phases that remains relatively unstretched.

\vspace{2mm}
\noindent Symmetry-breaking phase transformations involve changes in lattice geometries, and under certain conditions, these changes can result in a \textit{volume-preserving} deformation. In such materials, the stretch tensor $\mathbf{U}$ satisfies the condition $|\mathrm{det}\mathbf{U} - 1|=0$. The determinant of $\mathbf{U}$ represents the volume ratio between the transformed and reference lattices, and $|\mathrm{det}\mathbf{U} - 1|=0$ indicates that the transformation occurs without a net volume change of the unit cells. This condition is one of the important criteria necessary to form self-accommodating microstructures in shape-memory alloys. 

\vspace{2mm}
\noindent The compatibility conditions, namely the $\lambda_2 = 1$ condition to form stress-free interfaces and the $|\mathrm{det}\mathbf{U}-1|=0$ condition for volume-preserving microstructures, are more general than the zero-strain condition that is widely used in designing intercalation compounds \cite{schofield2022doping,chang2015elucidating,kim2011enabling,santos2023chemistry}. For example, both the $\lambda_2 = 1$ and $|\mathrm{det}\mathbf{U}-1|=0$ conditions can be satisfied by \textit{multiple} combinations of lattice geometries and symmetry-breaking transformation pathways. Fig.~\ref{Fig7}(a) shows several combinations of lattice geometries undergoing a cubic-to-tetragonal transformation that satisfy the $\lambda_2= 1$ and/or $|\mathrm{det}\mathbf{U}-1|=0$ conditions. These multiple combinations of lattice geometries provide several routes to engineer lattice deformations---beyond the zero-strain condition---by doping intercalation compounds, and thus offer greater flexibility to chemists in designing material compositions. Moreover, the zero-strain condition is a special case of the compatibility condition in Eq.~(\ref{eq: exact interface}), where the eigenvalues satisfy $\lambda_1 = \lambda_2 = \lambda_3 = 1$. This condition is met when the host lattices are rigid and do not distort, corresponding to the $\alpha = \beta = 1$ in Fig.~\ref{Fig7}(a), and results in zero volume change during phase transformation. By contrast, the solutions to Eq.~(\ref{eq: exact interface}) and $|\mathrm{det}\mathbf{U}-1|=0$ are more general and allow a wider range of lattice deformations satisfying $\lambda_1 \leq \lambda_2 = 1 \leq \lambda_3$. In our study, we use representative lattice geometries identified as `A' and `B' in Fig.~\ref{Fig7}(a) that satisfy the $\lambda_2 = 1$ and $\mathrm{det}\mathbf{U} = 1$ conditions, respectively. 

\vspace{2mm}
\noindent The crystallographic microstructures emerge phenomenologically in the material during phase transformations. In intercalation compounds, the electrochemical driving forces---in addition to elastic energy minimization---affect the pathways of microstructural evolution. For example, anisotropic Li-diffusion, thermodynamic and kinetic energy barriers, and galvanostatic/potentiostatic surface conditions, all affect microstructural evolution. These factors are not for accounted in the crystallographic theory of martensites (a sharp-interface theory), and the energy-minimizing sequences given by Eqs.~(\ref{eq:twin equation}), (\ref{eq:AM interface}) and (\ref{eq: exact interface}) are purely analytical solutions used to derive continuum microstructures. Therefore, it is important to analyze the nucleation and growth of the crystallographic microstructures within an electrochemical environment, and to investigate the collective energy barriers and driving forces that shape continuum microstructures.

\subsection*{Coupling Li-Diffusion and Host-Lattice Deformation}
\label{sec:Coupling Li-Diffusion and Host-Lattice Deformation}
\noindent We present a continuum theory for intercalation compounds undergoing symmetry-breaking phase transformations. Unlike existing phase-field models \cite{tang2011anisotropic, nadkarni2019modeling, ombrini2023thermodynamics, di2014cahn}, which use Li-composition as the primary order parameter and typically homogenize and linearize lattice deformations, our approach accounts for individual lattice variants that emerge during symmetry-breaking transformations. We construct a multi-well free energy landscape as a function of both Li-composition $c$ and host-lattice deformation gradient $\mathbf{F}$. By minimizing the total energy across this landscape with flux boundary conditions, our continuum model predicts the nucleation and growth of crystallographic microstructures. This symmetry-related continuum theory is general and applicable to all symmetry-breaking intercalation compounds.\footnote{compounds in which host-lattices undergo displacive transformations.} For a detailed derivation and formulation of our continuum model, please refer to our previous work \cite{zhang2024coupling}.

\vspace{2mm}
\noindent
We begin by formulating the continuum theory in three dimensions and then reduce it to a two-dimensional framework. This dimensional reduction is achieved by assuming in-plane lattice deformations and selecting appropriate 2D projections based on the orientations of compatible interfaces. The well-characterized compound Li$_2$Mn$_2$O$_4$ serves as a representative material for this study. Our findings provide quantitative insight into the internal stresses and volume changes that accompany phase transformations in Li$_2$Mn$_2$O$_4$. Moreover, our multi-variant continuum model serves as a computational tool for crystallographically designing stress-free interfaces and volume-preserving microstructures in intercalation compounds. These theoretical results potentially suggest new ways for designing intercalation compounds with enhanced reversibility.

\vspace{2mm}
\noindent In our computations, we model a domain of length $\mathrm{L}= 500~\mathrm{nm}$, with periodic boundary conditions applied to the surfaces $x_1 = 0$ and $x_1 = \mathrm{L}$ (and a thickness of 40~nm for three-dimensional calculations). A Li-flux is applied on all surfaces at a constant 0.5C-rate, and we solve for mechanical equilibrium at each step of the microstructural evolution. In all calculations presented in this work, we initialize the computational domain in the uniform reference LiMn$_2$O$_4$ phase.

\subsubsection*{Free Energy}
We construct a multi-well free-energy landscape for Li$_2$Mn$_2$O$_4$ using two order parameters, namely the normalized Li-composition $\bar{c}$ and a strain measurement vector $\mathbf{e} = \{e_1, \ e_2,\ e_3,\ \dots, \ e_6\}^\top$. The Li-composition distinguishes between the reference and intercalated phases, while the components of the strain measurement vector (e.g., $e_2$, $e_3$) distinguish between lattice variants in the transformed phase. The strain measurements are linear combinations of the components of the Green-Lagrange strain tensor $\mathbf{E}$, and are given by \cite{barsch1984twin}:
\begin{align}
e_1 &=\frac{1}{\sqrt {3}}(E_{11}+E_{22}+E_{33}),\nonumber \\ 
e_2 &=\frac{1}{\sqrt {2}}(E_{11}-E_{22}),\nonumber \\ 
e_3 &=\frac{1}{\sqrt {6}}(E_{11}+E_{22}-2E_{33}),\nonumber \\ 
e_4 &=\sqrt {2}E_{23} = \sqrt {2}E_{32},\nonumber \\ 
e_5 &=\sqrt {2}E_{13} = \sqrt {2}E_{31},\nonumber \\ 
e_6 &=\sqrt {2}E_{12} = \sqrt {2}E_{21}.
\label{eq:3D_strain measure}
\end{align}

\noindent The Green-Lagrange strain is a symmetric tensor related to the host-lattice deformation gradient as $\mathbf{E}=\frac{1}{2}(\mathbf{F}^{\top}\mathbf{F}-\mathbf{I})$.

\vspace{2mm}
\noindent We formulate the total free-energy density of an intercalation compound that undergoes a symmetry-breaking transformation in 3D as: 
\begin{align}
     \Psi(\mathbf e, \nabla \mathbf e,  \bar{c}, \nabla \bar{c})&=\int_{\Omega}\underbrace{\mathrm{RT_0c_0}\biggl(\frac{T}{\mathrm{T_0}}\left[\bar{c}\operatorname{ln} \bar{c}+\left(1-\bar{c}\right)\operatorname{ln}\left(1-\bar{c}\right)\right]+\mu_0 \bar{c}+\bar{c}(1-\bar{c})\sum_{i=1}^{n}\alpha_{i}(1-2\bar{c})^{i-1}\biggr)}_\text{$\psi_{\mathrm{ther}}(\bar c)$}\nonumber\\
     &+\frac{3}{2}\mathrm{K}(e_1 - \frac{\bar{c} - 0.5}{1.0 - 0.5} \Delta\mathrm{V})^2+ \mathrm{G}(e_4^2 + e_5^2 + e_6^2)\nonumber\\
     &+\underbrace{\mathrm{C}\frac{\bar{c}-0.75}{0.5-0.75} (e_2^2 + e_3^2)+ \beta_0 \frac{\bar{c} - 0.75}{1.0 - 0.75} e_3 (e_3^2 - 3e_2^2) + \beta_1 (e_2^2 + e_3^2)^2}_\text{$\psi_{\mathrm{elas}}(\mathbf e, \bar{c})$} \nonumber\\
    &+\underbrace{\frac{\mathrm{RT_0c_0}}{2}(\nabla \bar c \cdot \lambda \nabla \bar c + \nabla e_{2} \cdot \kappa \nabla  e_{2}+\nabla e_3 \cdot \theta \nabla  e_3)}_\text{$\psi_{\mathrm{grad}} (\nabla \bar c, \nabla e_2, \nabla e_3)$}~\mathrm{d}{\mathbf{x}}.
    \label{eq:3D_free_energy}
\end{align}

\noindent Fig.~\ref{Fig4}(a) shows a three-dimensional plot of the free energy, corresponding to Eq.~(\ref{eq:3D_free_energy}), as a function of the strain measure components $e_2$, $e_3$ and the normalized concentration $\bar c$. At $(e_2,~e_3,~\bar{c}) = (0,~0,~0.5)$, there exists an energy well corresponding to the higher-symmetry cubic phase. This cubic phase is represented by the identity tensor~$\mathbf{I}$. Three additional energy wells are located at $(e_2,~e_3,~\bar{c})$ with $(0.1,~0.07,~1.0)$, $(-0.1,~0.07,~1.0)$, and $(0,~-0.11,~1.0)$. These wells correspond to the tetragonal variants in the lower symmetry phase. The tetragonal variants are represented by the stretch tensors~$\mathbf{U}_1$, $\mathbf{U}_2$ and $\mathbf{U}_3$, respectively. We note that the free energy density in Eq.~(\ref{eq:3D_free_energy}) satisfies both frame-indifference and material symmetry conditions $\psi(\mathbf{Re},\mathbf{R}\nabla(\mathbf{e})\mathbf{R}^{\top},c,\mathbf{R}\nabla c) = \psi(\mathbf{e},\nabla\mathbf{e},c,\nabla c)$ for all rotations $\mathbf{R}$ in the finite point group of the undistorted crystalline lattice $\mathcal{P}(\mathbf{e}^\circ_i)$ \cite{zhang2024coupling}.

\vspace{2mm}
\noindent We assume a two-dimensional form of the model with $E_{13} = E_{23} = E_{33} = 0$, and reduce the strain measures in Eq.~(\ref{eq:3D_strain measure}) to:
$e_1 = \frac{1}{\sqrt {2}}(E_{11}+E_{22})$,
$e_2 = \frac{1}{\sqrt {2}}(E_{11}-E_{22})$, and
$e_6 = \sqrt {2}E_{12} = \sqrt {2}E_{21}$. Using these strain measurements, we construct the free energy in 2D as a function of $e_1$, $e_2$, and $e_6$:
% with $e_4 = e_5 = 0$ and $e_3 = e_1/\sqrt{2}$. Using these strain measurements, we construct the free energy in 2D as a function of $e_1$, $e_2$, and $e_6$:
\begin{align}
\Psi(\mathbf e, \nabla \mathbf e,  \bar{c}, \nabla \bar{c})&=\int_{\Omega}\underbrace{\mathrm{RT_0c_0}\biggl(\frac{T}{\mathrm{T_0}}\left[\bar{c}\operatorname{ln} \bar{c}+\left(1-\bar{c}\right)\operatorname{ln}\left(1-\bar{c}\right)\right]+\mu_0 \bar{c}+\bar{c}(1-\bar{c})\sum_{i=1}^{n}\alpha_{i}(1-2\bar{c})^{i-1}\biggr)}_\text{$\psi_{\mathrm{ther}}(\bar c)$}\nonumber\\
&+\underbrace{\mathrm{K}(e_1 - \frac{\bar{c}-0.5}{1.0-0.5}\Delta \mathrm{V})^2 + \mathrm{C}\frac{\bar{c}-0.75}{0.5-0.75}e_2^2 + \beta_1 e_2^4 + \mathrm{G}e_6^2}_\text{$\psi_{\mathrm{elas}}(\mathbf e, \bar{c})$}\nonumber\\
&+\underbrace{\frac{\mathrm{RT_0c_0}}{2}(\nabla \bar c \cdot \lambda \nabla \bar c + \nabla e_{2} \cdot \kappa \nabla  e_{2})}_\text{$\psi_{\mathrm{grad}} (\nabla \bar c, \nabla e_2)$}~\mathrm{d}{\mathbf{x}}.
\label{eq:free energy}
\end{align}

\vspace{2mm}
\noindent Eq.~(\ref{eq:free energy}) describes a multi-well free-energy landscape that not only differentiates between the reference and intercalated phases---located at $\bar{c}=0.5$ and $\bar{c}=1.0$, respectively---but also distinguishes between the two tetragonal lattice variants at $e_2=\pm 0.1$ and the cubic lattice at $e_2=0$. Fig.~\ref{Fig4}(b) shows this multi-well energy landscape, calibrated for Li$_{1-2}$Mn$_2$O$_4$, in two dimensions.

\vspace{2mm}
\noindent The total free energy accounts for thermodynamic, elastic, and gradient energy contributions. In Eqs.~(\ref{eq:3D_free_energy}) and (\ref{eq:free energy}), we use the Redlich-Kister polynomial series to construct the thermodynamic energy of the intercalation compound \cite{redlich1948algebraic}, and we fit the coefficients $\mu_0$ and $\alpha_i$ to the open-circuit voltage (OCV) curve of Li$_2$Mn$_2$O$_4$ \cite{thackeray1983lithium}. The Legendre transformation of $\psi_{\mathrm{ther}}(\bar{c})$ results in a double-well potential with minima at $\bar{c}=0.5$ and $\bar{c}=0.99$ (see Fig.~\ref{Fig4}(b)):
\begin{align}
\tilde\psi_{\mathrm{ther}}(\bar c) =\psi_{\mathrm{ther}}(\bar c)-\frac{\partial \psi_{\mathrm{ther}}(\bar c=0.99)}{\partial \bar c}\bar c.
\label{eq:Legendre transform}
\end{align}

\noindent In Eq.~(\ref{eq:free energy}), the polynomial constructed using the strain measurement order parameters ($e_1$, $e_2$, and $e_6$) describes a multi-well energy landscape that distinguishes between the individual lattice variants resulting from the symmetry-breaking phase transformation. For small deformations, the $e_1$ parameter governs dilatational changes of the cubic phase, the $e_2$ parameter breaks the 90$^\circ$ rotational symmetry of the cubic phase, generating the multi-well structure characteristic of the tetragonal phase, and the $e_6$ parameter controls shearing of the cubic phase. The bulk modulus $\mathrm{K}$, shear modulus $\mathrm{G}$, and deviatoric modulus $\mathrm{C}$ correspond to the elastic constants of LiMn$_2$O$_4$, and are fitted as $\mathrm{K}=\frac{c_{11}+c_{12}}{2}$, $\mathrm{G}=c_{44}$, and $\mathrm{C}=\frac{c_{11}-c_{12}}{2}$, respectively. The coefficients $\beta_0$, $\beta_1$ and the constant $\Delta \mathrm{V}$ are calibrated to satisfy the equilibrium condition:
\begin{align}
    \left.\frac{\partial\psi_{\mathrm{elas}}(\mathbf e, \bar{c})}{\partial e_i}\right\vert_{\bar{c}=1.0,~\mathbf{e}=\mathbf{e}_0} = 0.
    \label{eq: equilibrium strain measurement}
\end{align}

\noindent In Eq.~(\ref{eq: equilibrium strain measurement}) $\mathbf{e}_0$ corresponds to the spontaneous strain values for Li$_2$Mn$_2$O$_4$ derived in the Stretch Tensor section.

\vspace{2mm}
\noindent Finally, we model isotropic gradient energy terms to penalize changes in Li-composition and strain. The corresponding gradient energy coefficients, $\lambda$, $\kappa$ and $\theta$, are numerically calibrated to describe the diffuse phase boundaries between the reference and transformed phases, and the compatible twin boundaries between lattice variants, respectively.

\subsubsection*{Governing Equations}
We compute the evolution of Li-composition following the law of mass conservation:
\begin{align}
    \frac{\partial c}{\partial t}+\nabla \cdot \mathbf{j}=0.\label{eq:mass balance}
\end{align}

\noindent In Eq.~(\ref{eq:mass balance}), $\mathbf{j}$ is the diffusive flux, defined by the Onsager relation $\mathbf{j}=-\mathbf{M}(c)\nabla\mu$. This flux depends on the mobility tensor $\mathbf{M}(c)$ and the gradient of the chemical potential $\mu$. Li-diffusion in $\mathrm{Li_2Mn_2O_4}$ is isotropic \cite{erichsen2020tracking}, and we formulate the mobility tensor as $\mathbf{M}(c) = \frac{\mathrm{D_0}c(\mathrm{c_0} - c)}{\mathrm{RT_0} c_0} \mathbf{I}$, in which $\mathrm{D}_0$ is the Li-diffusion coefficient. The chemical potential $\mu$ is defined as the variational derivative of the free energy:
\begin{align}
    \mu=\frac{1}{c_0}\frac{\partial\psi_{\mathrm{ther}}}{\partial \bar{c}}+\frac{1}{c_0} \frac{\partial \psi_{\mathrm{elas}}}{\partial \bar{c}}- \mathrm{RT_0}(\nabla \cdot \lambda \nabla \bar{c}).
    \label{eq:total chemical potential}
\end{align}

\noindent The individual terms in Eq.~(\ref{eq:total chemical potential}) represent the thermodynamic, elastic, and gradient energy contributions to the chemical potential. These contributions vary across space and time, and we analyze their effects on phase transformation kinetics in the Results section. In particular, we find that the elastic-energy-driven component of the chemical potential contributes anisotropically to Li-diffusion along and across twin boundaries---a feature not previously captured in phase-field models for intercalation compounds.

\vspace{2mm}
\noindent We neglect surface wetting on the electrode particle (i.e., $\nabla c \cdot \hat{\mathbf{n}}=0$) and model a constant Li-flux on all surfaces as:
\begin{align}
    \mathbf{j}\cdot\hat{\mathbf{n}}=-\frac{\mathcal{C}\mathrm{c_0L}}{3600}.
    \label{eq: mass flux}
\end{align}

\noindent In Eq.~(\ref{eq: mass flux}) $\mathcal{C}$ represents the C-rate and $\mathrm{L}$ corresponds to the computational domain size.

\vspace{2mm}
\noindent In our calculations, we assume that mechanical relaxation occurs instantaneously compared to the timescale of Li-diffusion. In the absence of external mechanical loads on the electrode, we solve for the mechanical equilibrium conditions as:
\begin{align}
    \nabla \cdot \mathbf{T}_{\mathrm{R}}^{\top}-\nabla \cdot(\nabla \cdot \mathbf{Y}^{\top})^{\top}=0.
    \label{eq:macroscopic force balance}
\end{align}
In Eq.~(\ref{eq:macroscopic force balance}) $\mathbf{T}_{\mathrm{R}}$ and $\mathbf{Y}$ denote the first Piola-Kirchhoff stress tensor and the higher-order stress tensor, respectively. These stress tensors are related to the variational derivatives of the free energy as:
\begin{align}
    \mathbf{T}_{\mathrm{R}}&= \sum_{i}\left(\frac{\partial \psi_{\mathrm{elas}}}{\partial e_i} \frac{\partial e_i}{\partial \mathbf{F}} + \frac{\partial \psi_{\mathrm{grad}}}{\partial \nabla e_{i}} \frac{\partial\nabla e_{i}}{\partial \mathbf{F}}\right),\\
    \mathbf{Y}&=  \sum_{i}\frac{\partial \psi_{\mathrm{grad}}}{\partial\nabla e_{i}} \frac{\partial\nabla e_{i}}{\partial \nabla\mathbf{F}}.
\end{align}

\noindent The first Piola-Kirchhoff stress tensor $\mathbf{T}_\mathrm{R}$ is defined in the reference configuration. For our analysis, we convert it to the classical Cauchy stress tensor $\boldsymbol{\sigma}$, which is defined in the deformed configuration, as follows:
\begin{align}
    \mathbf{T}_\mathrm{R} = J \boldsymbol{\sigma} \mathbf{F}^{-\top}.
    \label{eq:pk1-cauchy}
\end{align}

\noindent In Eq.~(\ref{eq:pk1-cauchy}), the Jacobian determinant is given by $J = \mathrm{det}\mathbf{F}$. Detailed derivations of the constitutive equations for Li-diffusion and the finite deformation of host lattices are described in our recent work \cite{zhang2024coupling}. We implement our multi-variant continuum model within a finite-element framework. Specifically, we use the Galerkin weak forms of the dimensionless Eqs.~(\ref{eq:mass balance}), (\ref{eq:total chemical potential}) and (\ref{eq:macroscopic force balance}) in the open-source, parallel finite-element framework (MOOSE). We solve the system of nonlinear equations using the preconditioned Jacobian-free Newton-Krylov method, and compute time integration using the implicit Backward-Euler method. Further details on the numerical implementation of our model are provided in Ref.~\cite{zhang2024coupling} and outlined in Appendix~C.

\vspace{2mm}
\noindent {\bf Data Availability}. The authors declare that the data supporting the findings of this study are available in the paper and its supplementary information files. %All codes developed in this study are open-source and available on the OSF repository at \href{https://osf.io/45bqh/?view_only=6716a84098814cb68347a4044e5fe02c}{OSF|Solids \& Materials Group (UCSB)}.

\vspace{2mm}
\noindent {\bf Acknowledgments}. The authors acknowledge research funding from the U.S. Department of Energy (DOE), Office of Basic Energy Sciences, Division of Materials Sciences and Engineering under Award DE-SC0024227 (Model development; DZ, ARB). ARB acknowledges the support of the Air Force Fiscal Year 2023 Young Investigator Research Program, U.S. under Grant No. FA9550-23-1-0233 (symmetry calculations; ARB). The authors thank the Center for Scientific Computing at the University of California, Santa Barbara (MRSEC; NSF DMR 2308708) for providing computational resources that contributed to the results reported in this paper.

\vspace{2mm}
\noindent {\bf Author contribution}. ARB conceptualized the project and obtained funding. DZ and ARB designed the methodology and worked together on model development, theoretical analysis and calculations, as well as data visualization. Both authors contributed to the writing of the paper.

\vspace{2mm}
\noindent {\bf Competing Interests}. The authors declare that they have no competing interests.

\clearpage
\printbibliography

\clearpage
\begin{appendices}     
\appendix
\setcounter{figure}{0}
\setcounter{equation}{0}
\renewcommand{\thesection}{Appendix \Alph{section}}
\renewcommand{\thetable}{S\arabic{table}}
\renewcommand{\thefigure}{S\arabic{figure}}
\renewcommand{\theequation}{\Alph{section}.\arabic{equation}}
\section{Symbols} \label{A}
We summarize all symbols used in our work in Table~\ref{tab:1}.

\begin{longtable}{p{2.4cm}p{8.5cm}p{2.4cm}}
    \hline
    Symbol&Description&Unit\\
    \hline
    $\Omega$&The reference body&[/]\\
    $\partial \Omega$&Surface of the reference body&[/]\\
    $c$&Species concentration in the reference configuration&[$\mathrm{mol/m^3}$]\\
    $\mathrm{c_0}$&Maximum reference species concentration &[$\mathrm{mol/m^3}$]\\
    $\bar{c}$&Normalized reference concentration ($\bar{c}=c/\mathrm{c_0}$)&[/]\\
    $\mathrm{L}$&Characteristic length&[$\mathrm{m}$]\\
    $t$&Time&[$\mathrm{s}$]\\
    % $\bar{t}$&Dimensionless time&[/]\\
    $\mathrm{D_0}$&Diffusion coefficient&[$\mathrm{m^2/s}$]\\
    $\mathrm{R}$&Gas constant&[$\mathrm{J/(mol\cdot K})$]\\
    $T$, $\mathrm{T_0}$&Temperature, Reference temperature (298 K)&[$\mathrm{K}$]\\
    $c_{11}/c_{12}/c_{44}$&Elastic constants&[Pa]\\
    K, G, C&Bulk modulus, Shear modulus, Deviatoric modulus&[Pa]\\
    $\mathbf{E}$&Green-Lagrange strain with components $E_{ij}$&[/]\\
    $\Delta \mathrm{V}$& Volume change during phase transformation &[/]  \\
    $\beta_0, \beta_1$&Elastic constants calibrated to satisfy the equilibrium condition in Eq.~(11)&[Pa]  \\
    $f$&Volume fraction&[/]\\
    $\mu_0, \alpha_i$&Thermodynamic properties&[/]\\
    $\mu$&Total chemical potential&[$\mathrm{J/mol}$]\\
    $\mu_\mathrm{elas}$&Stress chemical potential&[$\mathrm{J/mol}$]\\
    $\mu_{\mathrm{avg}}$&Average chemical potential&[$\mathrm{J/mol}$]\\
    % $\psi_{\mathrm{bulk}}$&Bulk free energy density&[$\mathrm{J/m^3}$]\\
    $\psi_{\mathrm{ther}}$&Thermodynamic free energy density&[$\mathrm{J/m^3}$]\\
    $\psi_{\mathrm{grad}}$&Gradient energy density&[$\mathrm{J/m^3}$]\\
    $\psi_{\mathrm{elas}}$&Elastic energy density&[$\mathrm{J/m^3}$]\\
    $\Psi_{\mathrm{avg}}$&Average free energy&[$\mathrm{J/m^3}$]\\
    % $\psi_{\mathrm{coup}}$&Coupled energy density&[$\mathrm{J/m^3}$]\\
    $\lambda, \kappa,\theta$&Gradient energy coefficient&[$\mathrm{m^2}$]\\
    $\mathbf{e}$&Strain measurements $\mathbf{e} = \{e_1, e_2, \dots, e_6\}^{\top}$&[/]\\
    $\mathbf{e}_0$&Spontaneous strain measurements&[/]\\
    $\nabla$ & Gradient operator&[$1/\mathrm{m}$]\\
    % $\bar{\nabla}$&Dimensionless nabla operator ($\nabla=\bar{\nabla}/\mathrm{L}$)&[/]\\
    $\mathbf{x}$ &Material points&[$\mathrm{m}$]\\
    % $\boldsymbol{\chi}$ &Mapping from material to spatial frame&[$\mathrm{m}$]\\
    $\mathbf{u}$&Displacement&[$\mathrm{m}$]\\
    $\mathbf{j}$&Species flux in the reference configuration&[$\mathrm{mol/(m^2\cdot s})$]\\
    % $\bar{\mathbf{j}}$&Dimensionless species flux ($\mathbf{j}=\frac{\mathrm{D_0c_0}}{\mathrm{L}}\bar{\mathbf{j}}$)&[/]\\
    % $\mathbf{t}$&Surface traction &[Pa]\\
    $\mathbf{\hat{n}}/ \mathbf{\hat{m}}/ \mathbf{a}/ \mathbf{b}$&Vector&[/]\\
    $\mathbf{F}$&Deformation gradient&[/]\\
    $\mathbf{U}_i/\mathbf{U}_j$& Stretch tensor&[/]\\
    $\mathbf{Q}/\mathbf{Q}'$&Rotation tensor&[/]\\
    $K$&Twin plane direction&[/]\\
    $\mathbf{T}_{\mathrm{R}}$&First Piola-Kirchhoff stress tensor&[Pa]\\
    $\mathbf{Y}$&Third-order stress tensor&[$\mathrm{Pa\cdot m}$]\\
    % $\boldsymbol{\rho}$&Lagrange multipliers &[Pa]\\
    $\mathbf{M}$&Mobility tensor &[$\mathrm{mol^2/(m\cdot J\cdot s)}$]\\
    % $\bar{\vct{M}}(c)$&Dimensionless mobility tensor ($\mathbf{M}=\frac{\mathrm{D_0c_0}}{\mathrm{RT_0}}\bar{\mathbf{M}}$)&[/]\\
    \hline
    \caption{A summary of symbols used in our work.}
    \label{tab:1}
\end{longtable}

\section{Structural data and geometric solutions} \label{B}

In this section, we list the structural data and geometric solutions of the crystallographic microstructures (e.g., stretch tensors, twin solutions, and volume fractions) presented in Fig.~2 of the main paper. For our study, we used representative intercalation compounds that have been characterized experimentally in the literature, including Li$_{2x}$Mn$_2$O$_4$ ($0.5 \leq x \leq 1$) \cite{erichsen2020tracking}, Na$_x$MnO$_2$ ($0 \leq x \leq 1$) \cite{abakumov2014multiple}, and Li$_x$[Li$_{1/3}$Ti$_{5/3}$]O$_4$ ($1 \leq x \leq 2$) \cite{ohzuku1995zero}.

\begin{table}[ht!]
    % \small
    \centering
    \begin{adjustbox}{width=1\textwidth}
    \renewcommand\arraystretch{1.5}
    % \addtolength{\leftskip} {-2cm}
    % \addtolength{\rightskip}{-2cm}
    \begin{tabular}{lll}
    \hline
    Stretch tensor& Twin solution & A/M solution\\
    \hline
    \\
    $\mathbf{U}_1 =\begin{bmatrix} 1.123&0&0\\0&0.969&0\\0&0&0.969\end{bmatrix}$& \begin{minipage}{.3\textwidth}
      \includegraphics[width=0.7\textwidth]{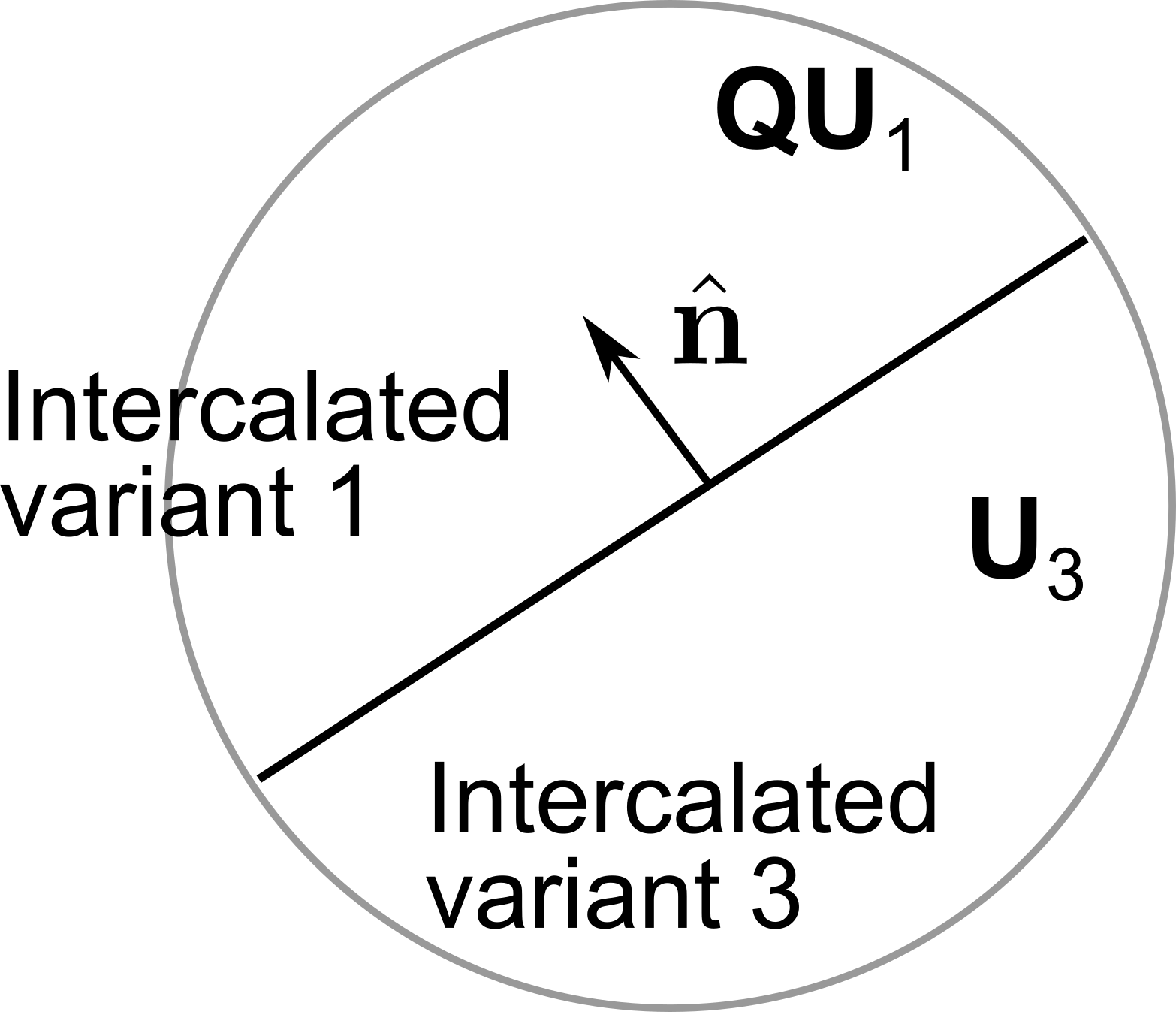}
    \end{minipage}&\begin{minipage}{.3\textwidth}
      \includegraphics[width=0.6\textwidth]{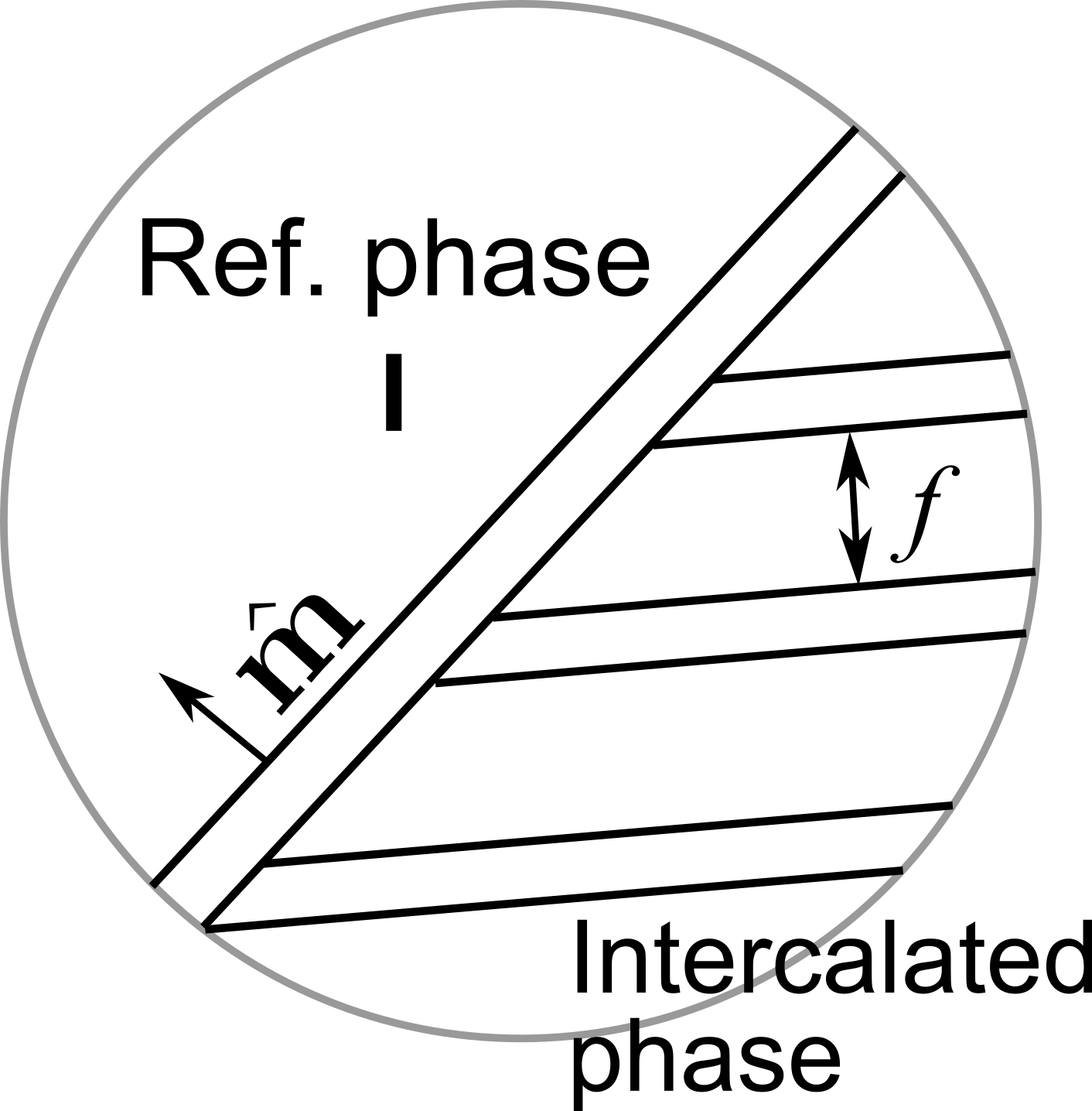}
    \end{minipage}\\
    \\
    \multirow{3}{15em}{$\mathbf{U}_2 =\begin{bmatrix} 0.969&0&0\\0&1.123&0\\0&0&0.969\end{bmatrix}$} & $\mathbf{a}=[\overline{0.200},\ 0,\ 0.232]$ & $f=0.216$\\
    &$\mathbf{\hat{n}}=(\overline{0.707},\ 0,\ \overline{0.707})$&$\mathbf{b}=[\overline{0.004}, \ 0.065, \ 0.010]$\\
    &$K=(\overline{0.757},\ 0,\ \overline{0.653})$&$\mathbf{\hat{m}}=(\overline{0.038}, \ \overline{0.500}, \ 0.865)$\\
    \\
    $\mathbf{U}_3 = \begin{bmatrix} 0.969&0&0\\0&0.969&0\\0&0&1.123\end{bmatrix}$ & $\mathbf{Q}= \begin{bmatrix} 0.989&0&0.146\\0&1&0\\-0.146&0&0.989\end{bmatrix}$ & $\mathbf{Q'}= \begin{bmatrix} 0.100&0.002&-0.032\\-0.001&0.999&0.051\\0.032&-0.051&0.998\end{bmatrix}$\\
    \\
    \hline
    \end{tabular}
    \end{adjustbox}
    \caption{The solutions to the twin interface given by Eq.~(1) and to the austenite/martensite interface given by Eq.~(2), respectively. We construct the twin interface using tetragonal variants $\mathbf{U}_1$ and $\mathbf{U}_3$ of Li$_2$Mn$_2$O$_4$. The solutions determine the orientation of twin planes (vectors $\mathbf{a}, \hat{\mathbf{n}}$), phase boundaries (vectors $\mathbf{b}, \hat{\mathbf{m}}$), rotation matrices $\mathbf{Q}$ and $\mathbf{Q}'$, and the volume fraction $f$ of twinned microstructures.}
    \label{Table2}
\end{table}

\vspace{2mm}
\noindent The cubic (LiMn$_2$O$_4$) to tetragonal (Li$_2$Mn$_2$O$_4$) symmetry-breaking transformation generates three variants, which we describe using stretch tensors $\mathbf{U}_1$, $\mathbf{U}_2$, and $\mathbf{U}_3$, see Table~\ref{Table2}. A twin interface can be described using any two of the three tetragonal variants. These twin interfaces, however, are rarely found in isolation, but instead form in large quantities as finely twinned mixtures that minimize the elastic energy at the LiMn$_2$O$_4$/Li$_2$Mn$_2$O$_4$ phase boundary. The average deformation of the finely twinned mixture is approximately coherent with the uniform cubic phase, and this two-phase microstructure is termed the austenite/martensite interface (a commonly used terminology in shape-memory alloys).

\vspace{2mm}
\noindent Table~\ref{Table2} lists the solutions for a representative twin interface and an austenite/martensite microstructure using variants $\mathbf{U}_1$ and $\mathbf{U}_3$. These solutions correspond to the analytically constructed microstructure shown in Fig.~2(a) of the main paper. These theoretical results show agreement with the experimental Bragg-filtered HRTEM measurements from Ref.~\cite{erichsen2020tracking}, which report a twin plane orientation of (1,~0,~1) and a volume fraction $f=0.2$.

\vspace{2mm}
\noindent Table~\ref{tab:NaMnO2} lists the structural data from Refs.~\cite{jain2013commentary,abakumov2014multiple} and twin solutions for microstructures in NaMnO$_2$. These solutions correspond to Fig.~2(c) of the main paper. Using these lattice geometries as input, we geometrically construct a twin interface $K = (0.705,\ 0,\ \overline{0.709})$ between two monoclinic variants, which aligns well with the HRTEM imaging of the twins along the $(1,~0,~\bar{1})$ direction reported by Abakumov et al. \cite{abakumov2014multiple}.
\begin{table}[ht!]
    \centering
    \renewcommand\arraystretch{1.5}
    \begin{tabular}{lll}
    \hline
    & $\mathrm{MnO_2}$ & $\mathrm{NaMnO_2}$ \\
    \hline
    Space group & $Pnnm$ & $C2/m$ \\
    Crystal symmetry & Orthorhombic & Monoclinic \\
    $\{a, b, c\}$ (\text{\AA}) & $\{2.86, 4.33, 4.46\}$ & $\{5.66, 2.86, 5.80\}$ \\
    $\{\alpha, \beta, \gamma\}$ ($^{\circ}$) & $\{90, 90, 90\}$ & $\{90, 113, 90\}$ \\
    \hline
    \multicolumn{3}{c}{Predicted Stretch Tensor} \\
    \hline
    \\
    \multicolumn{3}{c}{$\mathbf{U}_1=\begin{bmatrix} 1.281 & 0 & 0.260 \\ 0 & 1 & 0 \\ 0.260 & 0 & 1.274 \end{bmatrix}$ \quad
                         $\mathbf{U}_2=\begin{bmatrix} 1.281 & 0 & -0.260 \\ 0 & 1 & 0 \\ -0.260 & 0 & 1.274 \end{bmatrix}$} \\
    \\
    \hline
    Twin Solution & \multicolumn{2}{c}{$\mathbf{a} = [\overline{0.204},~0,~\overline{1}], \mathbf{\hat{n}} = (1,~0,~0), K = (0.705,~0,~\overline{0.709})$} \\
    \hline
    \end{tabular}
    \caption{Structural data of $\mathrm{MnO_2}$ and $\mathrm{NaMnO_2}$ from Ref.~\cite{jain2013commentary} and Ref.~\cite{abakumov2014multiple}, respectively. We use this data to calculate the stretch tensors $\mathbf{U}_1, \mathbf{U}_2$, and the geometric solutions for the twin interface.}
    \label{tab:NaMnO2}
\end{table}

\noindent Table~\ref{tab:4} lists the structural data accompanying the phase transformation between Li[Li$_{1/3}$Ti$_{5/3}$]O$_4$ and Li$_{2}$[Li$_{1/3}$Ti$_{5/3}$]O$_4$ \cite{ohzuku1995zero}. The table also lists our calculated values of the stretch tensor, the $\lambda_2$ value related to the stress-free interface, and the $|\mathrm{det}\mathbf{U}-1|$ value related to the volume-preserving (or self-accommodating) deformation. The negligible lattice mismatch at the phase boundary ($\lambda_2 \to 1$) contributes to the improved reversibility (over 100 cycles with 94$\%$ capacity retention) of the cathode compound \cite{ohzuku1995zero}.
\begin{table}[ht!]
    \centering
    \renewcommand\arraystretch{1.5}
    \begin{tabular}{llllll}
    \hline
    Compound& Symmetry change& Predicted stretch tensor&$\lambda_2 (\mathbf{I}/\mathbf{U})$ & $|\mathrm{det}\mathbf{U}-1|$\\
    \hline
    \\
    Li$_{1\rightarrow2}$[Li$_{1/3}$Ti$_{5/3}$]O$_4$& Cubic $\rightarrow$ Cubic& \multirow{2}{12em}{$\begin{bmatrix} 1.001&0&0\\0&1.001&0\\0&0&1.001 \end{bmatrix}$}&1.001&0.002\tabularnewline[20pt]\\
    \\
    \hline
    \end{tabular}
    \caption{Using the structural measurements reported in Ref.~\cite{ohzuku1995zero}, we calculate the stretch tensor $\mathbf{U}$, the middle eigenvalue $\lambda_2$, and the volume change $|\mathrm{det}\mathbf{U}-1|$ accompanying the Li[Li$_{1/3}$Ti$_{5/3}$]O$_4$ to Li$_{2}$[Li$_{1/3}$Ti$_{5/3}$]O$_4$ phase transformation.}
    \label{tab:4}
\end{table}

\section{Numerical implementation} \label{D}
\setcounter{equation}{0}
\noindent We outline the finite element implementation of our multi-variant continuum model. This model represents a coupled, nonlinear, initial/boundary value problem in which the Cahn-Hilliard type of diffusion and mechanical equilibrium equations (with nonlinear gradient elasticity) are solved. Further details of the model are described in Ref.~\cite{zhang2024coupling}.

\subsection{Mass balance and diffusion laws}
\vspace{2mm}
\noindent The total mass of the intercalating species (e.g., Li-ion) is conserved in accordance with the mass conservation law, and is given by:
\begin{align}
    \frac{\partial c}{\partial t}+\nabla \cdot \mathbf{j}=0.
    \label{eq:c1}
\end{align}

\noindent The flux $\mathbf{j}$ in Eq.~(\ref{eq:c1}) is related to the mobility tensor $\mathbf{M}(c)$ and the gradient of the chemical potential $\mu$ via the Onsager relation:
\begin{align}
    \mathbf{j}=-\mathbf{M}(c)\nabla \mu.
    \label{eq:c2}
\end{align}

\noindent By substituting Eq.~(\ref{eq:c2}) into Eq.~(\ref{eq:c1}), we obtain the mass balance law as:
\begin{align}
    \frac{\partial c}{\partial t}=\nabla \cdot (\mathbf{M}(c)\nabla \mu).
    \label{eq:c3}
\end{align}

\noindent We consider $\partial \Omega = \partial \Omega^{\{c\}} \cup \partial \Omega^{\{\mathbf{j}\}}$ to be complementary subsurfaces of the boundary, in which the species concentration is specified on $\partial \Omega^{\{c\}}$ in Eq.~(\ref{eq:CH Dirichlet}) and the global flux on $\partial \Omega^{\{\mathbf{j}\}}$ in Eq.~(\ref{eq:CH Neumann}). Additionally, the initial condition for the concentration field is given by Eq.~(\ref{eq:CH InitialBC}).
\begin{align}
    c=\breve{c} &\quad\text{on $\partial \Omega^{\{c\}}$},\label{eq:CH Dirichlet}\\
  \mathbf{j}\cdot \hat{\mathbf{n}} = -\frac{\mathcal{C}\mathrm{c_0L}}{3600} &\quad\text{on $\partial \Omega^{\{\mathbf{j}\}}$},\label{eq:CH Neumann}\\
  c ({\mathbf x},~t)= \tilde c ({\mathbf x}) &\quad\text{at $t=0$}.\label{eq:CH InitialBC}
\end{align}

\vspace{2mm}
\noindent We used a mixed formulation method in which $\mu$ and $c$ are treated as primary unknowns. By multiplying Eq.~(\ref{eq:c3}) and Eq.~(13) in the main paper by the variational test functions $\delta \bar c$ and $\delta \mu$, respectively, and applying the divergence theorem, we obtain the weak form of the Cahn-Hillard-type diffusion equation as follows:
\begin{align}
    \int_{\Omega}\frac{\partial c}{\partial t} \delta \mu~dV+  \int_{\Omega} \mathbf{M}\nabla \mu \cdot \nabla(\delta \mu)~dV-
    \int_{\partial \Omega}(\mathbf{M} \nabla \mu  \cdot \hat{\mathbf{n}} )\delta \mu ~dA=0,\\
    \int_{\Omega}\left(\frac{1}{\mathrm{c_0}}\frac{\partial( \psi_{\mathrm{ther}}+\psi_{\mathrm{elas}})}{\partial \bar{c}}-\mu\right)\delta \bar{c}~dV+\mathrm{RT_0}\int_{\Omega}\nabla(\delta \bar{c}) \cdot (\lambda\nabla \bar{c})~dV-\mathrm{RT_0}\int_{\partial\Omega}(\lambda \nabla \bar{c}\cdot \hat{\mathbf{n}})\delta \bar{c}~dA=0.
\end{align}

\subsection{Mechanical equilibrium}
\noindent The mechanical equilibrium condition corresponding to Eq.~(15) in the main paper can be written in indicial notation as:
\begin{align}
    T_{R_{iJ,J}}-Y_{iJK,JK} = 0.
    \label{eq:C1}
\end{align}
 
\noindent By introducing the Lagrange multipliers:
\begin{align}
    \rho_{iJ} = Y_{iJK,K},
    \label{eq:C8}
\end{align}
into Eq.~(\ref{eq:C1}), we have:
\begin{align}
T_{R_{iJ,J}}-\rho_{iJ,J}=0.
\label{eq:C9}
\end{align}

\noindent We solve for mechanical equilibrium by defining the kinematic relation between the deformation gradient $\mathbf{F}$ and the displacement field $\mathbf{u}$ as:
\begin{align}
F_{iJ}=u_{i,J}+\delta_{iJ},
\label{eq:C10}
\end{align}

\noindent in which, $\delta_{iJ}$ is the Kronecker delta ($\delta_{iJ}$ = 1 iff $i$ = $J$ and is zero otherwise). We numerically solve the weak forms of Eqs.~(\ref{eq:C8})-(\ref{eq:C10}) in a mixed-type finite element formulation and use suitable test functions ($\delta \mathbf{u}$, $\delta \mathbf{F}$ and $\delta \boldsymbol{\rho}$). By integrating Eqs.~(\ref{eq:C8})-(\ref{eq:C10}) over the reference volume $\Omega$, we obtain the following Galerkin weak forms:

\begin{gather}
    \int_{\Omega}(T_{R_{iJ,J}}-\rho_{iJ,J})\delta u_i  ~dV = 0, \\
    \int_{\Omega}(-Y_{iJK,K} + \rho_{iJ})  \delta F_{iJ} ~dV = 0, \\
    \int_{\Omega}(F_{iJ} - u_{i,J} - \delta_{iJ})  \delta \rho_{iJ} ~dV = 0.
\end{gather}

\subsection{Material constants}
\label{C3}
\noindent Table~\ref{Tab5} lists the material parameters used in our computations for the three representative materials: Li$_{1-2}$Mn$_2$O$_4$ and the two crystallographically designed compounds that respectively satisfy the stress-free ($\lambda_2 = 1$) interface and volume-preserving ($|\mathrm{det}\mathbf{U}-1|=0$) deformation.
\begin{table}[H]
    \centering
    \renewcommand\arraystretch{1.5}
    \begin{tabular}{llll}
    \hline
    &Parameter& Value&Source\\
    \hline
    Chemical&$\mu_0$&$-415.6$&Fitted to Ref.~\cite{thackeray1983lithium}\\
    &$\{\alpha_1$, $\alpha_2$, $\alpha_3\}$&$\{-597.2, \ -600.0, \ -304.2\}$&Fitted to Ref.~\cite{thackeray1983lithium}\\
    &$\lambda$&$7\times10^{-14} \ (\mathrm{m}^2)$&Ref.~\cite{zhang2024coupling}\\
    &$\kappa$, $\theta$&$7\times10^{-14} \ (\mathrm{m}^2)$&Ref.~\cite{zhang2024coupling}\\
    &$\mathrm{D_0}$&$2\times10^{-14} \ (\mathrm{m}^2/\mathrm{s})$&Ref.~\cite{christensen2006mathematical}\\
    &$\mathrm{c_0}$&$4.58\times10^{4} \ (\mathrm{mol}/\mathrm{m}^3)$&Ref.~\cite{zhang2007numerical}\\
    \hline
    Elastic&$c_{11}$&190.75 (GPa)&Ref.~\cite{zhang2024coupling}\\
    &$c_{12}$&36.63 (GPa)&Ref.~\cite{zhang2024coupling}\\
    &$c_{44}$&90.45 (GPa)&Ref.~\cite{zhang2024coupling}\\
    % &K&113.69 (GPa)&\\
    % &G&90.45 (GPa)&\\
    % &C&77.06 (GPa)&\\
    &$\Delta \mathrm{V}, \ \beta_1$(2D)&\multicolumn{2}{l}{Li$_2$Mn$_2$O$_4$: 0.07, 3098.82 (GPa)}\\
    &&\multicolumn{2}{l}{Stress-free interface: 0.06, 9232.82 (GPa)}\\
    &&\multicolumn{2}{l}{Volume-preserving deformation: 0.03, 3710.70 (GPa)}\\
    &$\Delta \mathrm{V}, \ \beta_0,\ \beta_1$(3D)&\multicolumn{2}{l}{Li$_2$Mn$_2$O$_4$: 0.04, 113.09 (GPa), 2997.36 (GPa)}\\
    \hline
    \end{tabular}
    \caption{Material parameters used in our computations for the three representative examples, namely Li$_{2x}$Mn$_2$O$_4$ (0.5 $\leq x \leq$ 1) and the two crystallographically designed systems satisfying the $\lambda_2=1$ and $|\mathrm{det}\mathbf{U}-1|=0$ conditions, respectively.}
    \label{Tab5}
\end{table}

\subsection{Analyzing the thermodynamic energy barrier}
\label{C4}
\begin{figure}[ht!]
    \centering
    \includegraphics[width=0.8\textwidth]{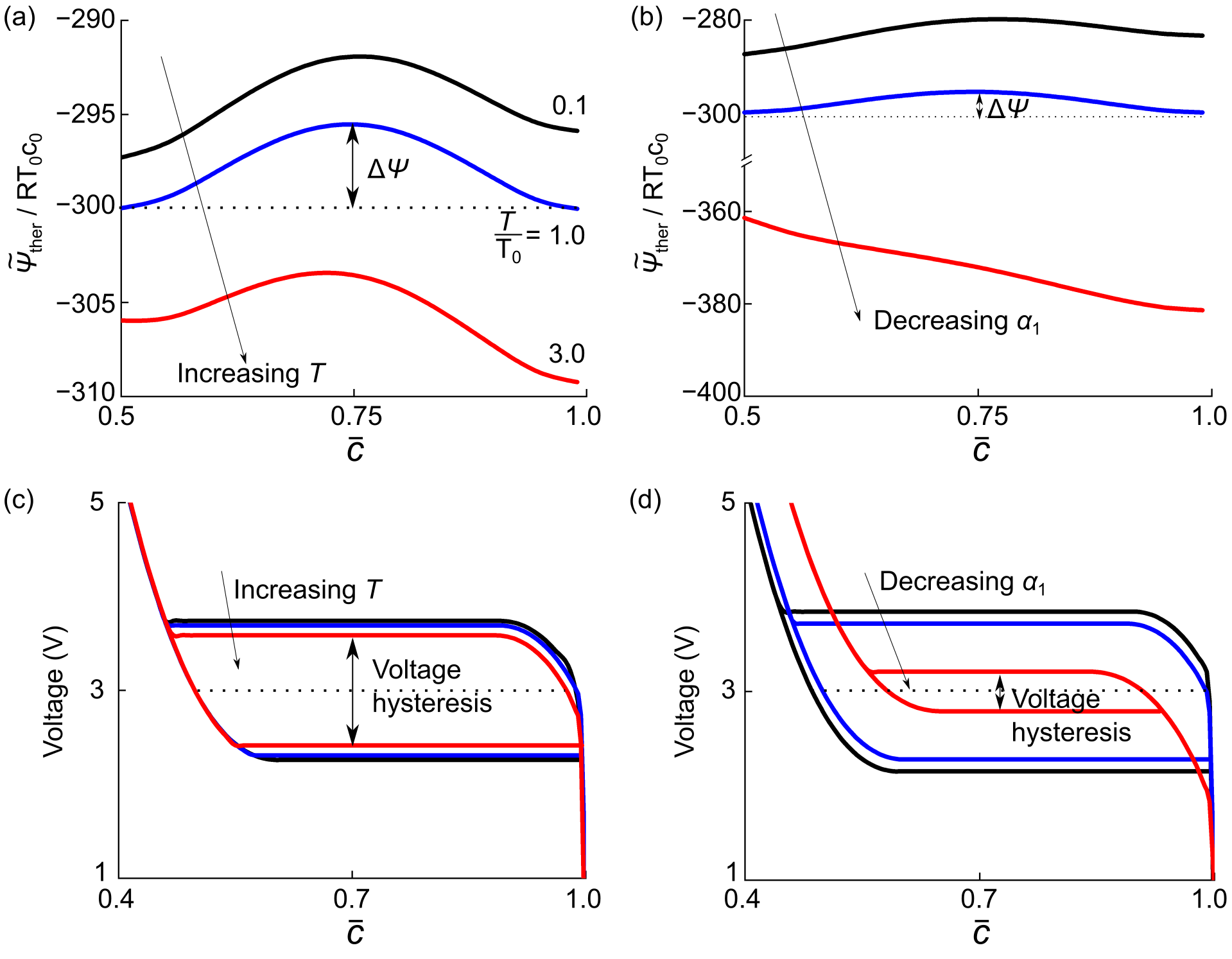}
    \caption{The effect of (a) temperature $T$ and (b) an enthalpy of mixing coefficient $\alpha_1$ on the thermodynamic energy barrier $\Delta\psi$. (a) The thermodynamic energy barrier $\Delta\psi$ decreases with increasing temperature $T$ ($\mathrm{T_0}$ = 298 K is the reference temperature). (b) Decreasing the value of the enthalpy of mixing coefficient $\alpha_1$ reduces the thermodynamic energy barrier $\Delta\psi$. (c, d) Voltage hysteresis loops corresponding to subfigures (a) and (b), respectively. Reducing the thermodynamic energy barrier $\Delta\psi$ further narrows the voltage hysteresis loops.}
    \label{FS2}
\end{figure}

\noindent The voltage hysteresis loops shown in Fig.~12(b) of the main paper are narrower by engineering the lattice compatibility between the two phases. Compatible phase boundaries, such as in materials that satisfy the $\lambda_2 = 1$ condition, offer near-zero interfacial stresses and are associated with reduced driving forces, which collectively reduce the width of the voltage hysteresis loop. However, these voltage hysteresis loops can be further reduced by minimizing the thermodynamic energy barriers. We explore this concept analytically in this section.

\vspace{2mm}
\noindent In our continuum model, we use the Redlich-Kister polynomial series ($n=3$) to construct the thermodynamic energy of the intercalation compound \cite{redlich1948algebraic}:
\begin{align}
    \psi_{\mathrm{ther}}(\bar c)=\mathrm{RT_0c_0}\biggl(\frac{T}{\mathrm{T_0}}\left[\bar{c}\operatorname{ln} \bar{c}+\left(1-\bar{c}\right)\operatorname{ln}\left(1-\bar{c}\right)\right]+\mu_0 \bar{c}+\bar{c}(1-\bar{c})\bigg[\alpha_1+\alpha_2(1-2\bar{c})+\alpha_3(1-2\bar{c})^2\bigg]\biggr).
    \label{eq:C16}
\end{align}

\vspace{2mm}
\noindent The Legendre transformation $\tilde\psi_{\mathrm{ther}}(\bar{c})$ of this thermodynamic energy density $\psi_{\mathrm{ther}}(\bar{c})$ is a double-well potential with minima at $\bar{c}=0.5$ and $\bar{c}=0.99$, see Fig.~3(b). The height of energy barrier in Eq.~(\ref{eq:C16}) is maximum at concentration $\bar{c}=0.75$, and we define this height as:
\begin{align}
    \Delta \psi&=2\tilde\psi_{\mathrm{ther}}(\bar c=0.75)-\tilde\psi_{\mathrm{ther}}(\bar c=0.5)-\tilde\psi_{\mathrm{ther}}(\bar c=0.99)\nonumber\\
    &=\mathrm{RT_0c_0}(-0.421\frac{T}{\mathrm{T_0}}+0.125\alpha_1-0.187\alpha_2+0.093\alpha_3).
    \label{eq:thermal barrier}
\end{align}

\noindent From Eq.~(\ref{eq:thermal barrier}) we note that the thermodynamic energy barrier $\Delta \psi$ is linearly proportional to the negative of the temperature (that is, $\Delta \psi \propto -T$), but also depends linearly on the enthalpy of the mixing coefficients ($\alpha_1, \alpha_2, \alpha_3$). In Fig.~\ref{FS2}(a) and Fig.~\ref{FS2}(b), we show representative examples of how increasing the temperature $T$ or decreasing $\alpha_1$, respectively, can reduce the height of the thermodynamic energy barrier. This reduction in the energy barriers contributes to the narrower hysteresis loops shown in Fig.~\ref{FS2}(c) and Fig.~\ref{FS2}(d), respectively. 

\subsection{Representative $\lambda_2=1$ computation}
\begin{figure}[ht!]
    \centering
    \includegraphics[width=\textwidth]{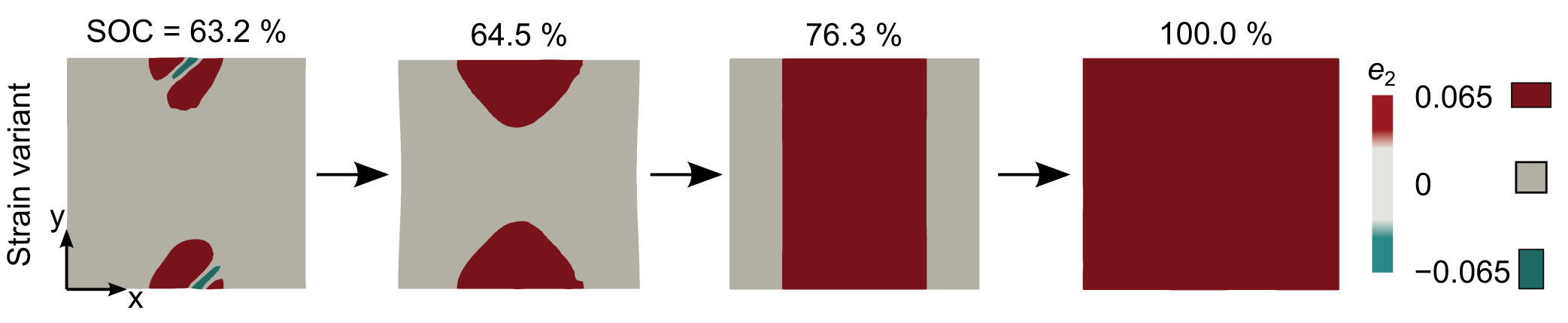}
    \caption{A representative calculation showing a stress-free interface satisfying the $\lambda_2 = 1$ condition nucleates and grows at 0.5C-rate. This evolution is similar to that shown in Fig.~8 of the main paper, albeit with a different tetragonal variant $e_2=0.065$.}
    \label{FS1}
\end{figure}
\noindent Fig.~\ref{FS1} shows the nucleation and growth of a microstructure satisfying the $|\lambda_2-1|=0$ condition. We note that the second type of tetragonal variant (i.e., $e_2=0.065$) emerges in this calculation and grows to form a compatible interface with the reference phase. This microstructure forms when periodic boundary conditions are applied to the top and bottom sides of the computational domain; this microstructure is equivalent to rotating the computational domain in Fig.~8 of the main paper by 90 degrees.

% \clearpage
% \printbibliography
% \bibliography{references}
\end{appendices}
\end{document}